\documentclass[rmp,aps,twocolumn,showpacs,amsmath,amssymb,floatfix]{revtex4-1}

\usepackage{multirow,amssymb,amsbsy,amsmath}
\usepackage{graphicx}
\usepackage{verbatim}
\def\bra#1{\mathinner{\langle{#1}|}}
\def\ket#1{\mathinner{|{#1}\rangle}}

\let \geq=\geqslant
\let \leq=\leqslant

\usepackage{color}

\begin{document}

\title{Non-Markovian dynamics in open quantum systems}

\author{Heinz-Peter Breuer}
\affiliation{Physikalisches Institut, Universit\"at Freiburg, 
Hermann-Herder-Stra{\ss}e 3, D-79104 Freiburg, Germany}

\author{Elsi-Mari Laine}
\affiliation{QCD Labs, COMP Centre of Excellence, Department of Applied 
Physics, Aalto University, 
P.O. Box 13500, 
FI-00076 AALTO, Finland}

\author{Jyrki Piilo}
\affiliation{Turku Centre for Quantum Physics, Department of Physics and
Astronomy, University of Turku, FI-20014 Turun yliopisto, Finland}

\author{Bassano Vacchini}
\affiliation{Dipartimento di Fisica, Universit\`a degli Studi di Milano,
Via Celoria 16, I-20133 Milan, Italy}
\affiliation{INFN, Sezione di Milano, Via Celoria 16, I-20133 Milan, Italy}

\date{\today}

\begin{abstract}
The dynamical behavior of open quantum systems plays a key role in many
applications of quantum mechanics, examples ranging from fundamental problems, 
such as the environment-induced decay of quantum coherence and relaxation in 
many-body systems, to applications in condensed matter theory, quantum transport, 
quantum chemistry and quantum information. In close analogy to a
classical Markovian stochastic process, the interaction of an open quantum system
with a noisy environment is often modeled phenomenologically by means of a dynamical 
semigroup with a corresponding time-independent generator in Lindblad form, which 
describes a memoryless dynamics of the open system typically leading to an irreversible
loss of characteristic quantum features. However, in many applications open 
systems exhibit pronounced memory effects and a revival of genuine quantum properties 
such as quantum coherence, correlations and entanglement. Here, recent theoretical 
results on the rich non-Markovian quantum dynamics of open
systems are discussed, paying particular attention to the rigorous mathematical definition, to
the physical interpretation and classification, as well as to the quantification of quantum 
memory effects. The general theory is illustrated by a series of physical examples. 
The analysis reveals that memory effects of the open system dynamics reflect characteristic 
features of the environment which opens a new perspective for applications, namely to 
exploit a small open system as a quantum probe signifying nontrivial features of
the environment it is interacting with. This article further explores the various physical 
sources of non-Markovian quantum dynamics, such as structured environmental spectral 
densities, nonlocal correlations between environmental degrees of freedom and 
correlations in the initial system-environment state, in addition to developing 
schemes for their local detection. Recent experiments addressing the detection, 
quantification and control of non-Markovian quantum dynamics are also briefly discussed.
\end{abstract}

\pacs{03.65.Yz, 03.65.Ta, 03.67.-a, 42.50.-p, 42.50.Lc}


\maketitle

\tableofcontents

\section{Introduction}

The observation and experimental control of characteristic quantum properties of
physical systems is often strongly hindered by the coupling of the system to a 
noisy environment. The unavoidable interaction of the quantum system with its
surroundings generates system-environment correlations leading to
an irretrievable loss of quantum coherence. Realistic quantum mechanical 
systems are thus open systems governed by a non-unitary time development 
which describes all features of irreversible dynamics such as the dissipation of 
energy, the relaxation to a thermal equilibrium or a stationary nonequilibrium state 
and the decay of quantum coherences and correlations
\cite{Davies1976,Alicki1987,Breuer2002}.

There is a well-established treatment of the dynamics of open quantum
systems in which the open system's time evolution is represented by
a dynamical semigroup and a corresponding master equation in Lindblad 
form \cite{Gorini1976a,Lindblad1976a}. 
If one adopts a microscopic system-environment approach to the dynamics of
open systems, such a master equation may be derived, for example, with the help 
of the Born-Markov approximation by assuming a weak coupling between the
system and its environment. However, it turns out that for many processes
in open quantum systems the approximations underlying this approach are not
satisfied and that a description of the dynamics by means of a dynamical 
semigroup fails. Typically, this is due to the fact that the relevant environmental 
correlation times are not small compared to the system's relaxation
or decoherence time, rendering impossible the standard Markov approximation.
The violation of this separation of time scales can occur, for example, in the
cases of strong system-environment couplings, structured or finite reservoirs, low 
temperatures, or large initial system-environment correlations.

If the dynamics of an open quantum system substantially deviates from that of a 
dynamical semigroup one often speaks of a non-Markovian process. This 
term refers to a well-known concept of the theory of classical stochastic 
processes and is used to loosely indicate the presence of memory effects in the 
time evolution of the open system. However, the classical notions of Markovianity 
and non-Markovianity cannot be transferred to the quantum regime in a natural 
way since they are based on a Kolmogorov hierarchy of joint probability 
distributions which does not exist in quantum theory 
\cite{Lindblad1979a,Accardi1982a,Vacchini2011a}. 
Therefore, the concept of a quantum non-Markovian process requires a precise 
definition which cannot be based on classical notions only. 
Many important questions need to be discussed in this context: How can 
one rigorously define non-Markovian dynamics in 
the quantum case, how do quantum memory effects manifest themselves in the 
behavior of open systems, and can such effects be uniquely identified 
experimentally through monitoring of the open system? The definition of 
non-Markovianity should provide a general mathematical characterization which 
does not rely on any specific representation or approximation of the dynamics, 
e.g. in terms of a perturbative master equation. Moreover, definitions of 
non-Markovianity should lead to quantitative measures for the degree of 
non-Markovianity, enabling the comparison of the amount of memory effects in 
different physical systems.

Recently, a series of different proposals to answer these questions has
been published, rigorously defining the border between the regions of
Markovian and non-Markovian quantum dynamics and developing
quantitative measures for the degree of memory effects (see, e.g., 
\cite{Wolf2008a,Breuer2009b,Rivas2010a,Luo2012a,Chruscinski2011a,Chruscinski2014a}, 
the tutorial paper \cite{Breuer2012a} and the recent review by \textcite{Rivas2014a}). 
Here, we describe and discuss several of these ideas, paying particular
attention to those concepts which are based on the exchange of
information between the open system and its environment, and on the
divisibility of the dynamical map describing the open system's time
evolution. We will also explain the relations between the classical
and the quantum notions of non-Markovianity and develop a general
classification of quantum dynamical maps which is based on these
concepts.

The general theory will be illustrated by a series of examples. We start
with simple prototypical models describing pure decoherence dynamics,
dissipative processes, relaxation through multiple decay channels and
the spin-boson problem. We then 
continue with the study of the dynamics of open systems which are coupled to 
interacting many-body environments. The examples include an Ising
and a Heisenberg chain in transverse magnetic fields, as well as an 
impurity atom in a Bose-Einstein condensate 
\cite{Apollaro2011a,Haikka2011a,Haikka2012a}. 
The discussion will demonstrate, in particular, that memory effects of the
open system dynamics reflect characteristic properties of the environment.
This fact opens a new perspective, namely to exploit a small open system 
as a quantum probe signifying nontrivial features of a complex environment, for 
example the critical point of a phase transition \cite{Smirne2013a,Gessner2014b}. 
Another example to be discussed 
here is the use of non-Markovian dynamics to determine nonlocal 
correlations within a composite environment, carrying out only 
measurements on the open system functioning as quantum 
probe \cite{Laine2012a,Wissmann2014a,Liu2013a}.

A large variety of further applications of quantum memory effects is described in the
literature. Interested readers may find examples dealing with, e.g., 
phenomenological master equations \cite{Mazzola2010a}, 
chaotic systems \cite{Znidaric2011a}, energy transfer processes in photosynthetic complexes
\cite{Rebentrost2011a}, 
continous variable quantum key distribution \cite{Vasile2011b}, metrology \cite{Chin2012a}, 
steady state entanglement \cite{Huelga2012a}, Coulomb crystals \cite{Borrelli2013a}, 
symmetry breaking \cite{Haas2013a}, and time-invariant quantum 
discord \cite{Haikka2013a}.

The standard description of the open system dynamics in terms of a 
dynamical map is based on the assumption of an initially factorizing 
system-environment state. The approach developed here also allows to 
investigate the impact of correlations in the initial system-environment state 
and leads to schemes for the local detection of such correlations through 
monitoring of the open system \cite{Laine2010b,Gessner2011a}. 

In recent years, several of the above features of non-Markovian 
quantum dynamics have been observed experimentally in photonic and trapped 
ion systems 
\cite{Liu2011a,Li2011a,Smirne2011a,Gessner2014a,Cialdi2014a,Tang2014}. 
We briefly discuss the results of these experiments which 
demonstrate the transition from Markovian to non-Markovian quantum dynamics, 
the non-Markovian behavior induced by nonlocal environmental correlations,
and the local scheme for the detection of nonclassical initial system-environment 
correlations.

\section{Definitions and measures for quantum non-Markovianity}
\label{Definitions-Measures}

\subsection{Basic concepts} \label{Basic-Concepts}

\subsubsection{Open quantum systems and dynamical maps}
\label{open-sys-dyn-maps}

\begin{figure}[tbh]
\includegraphics[width=0.48\textwidth]{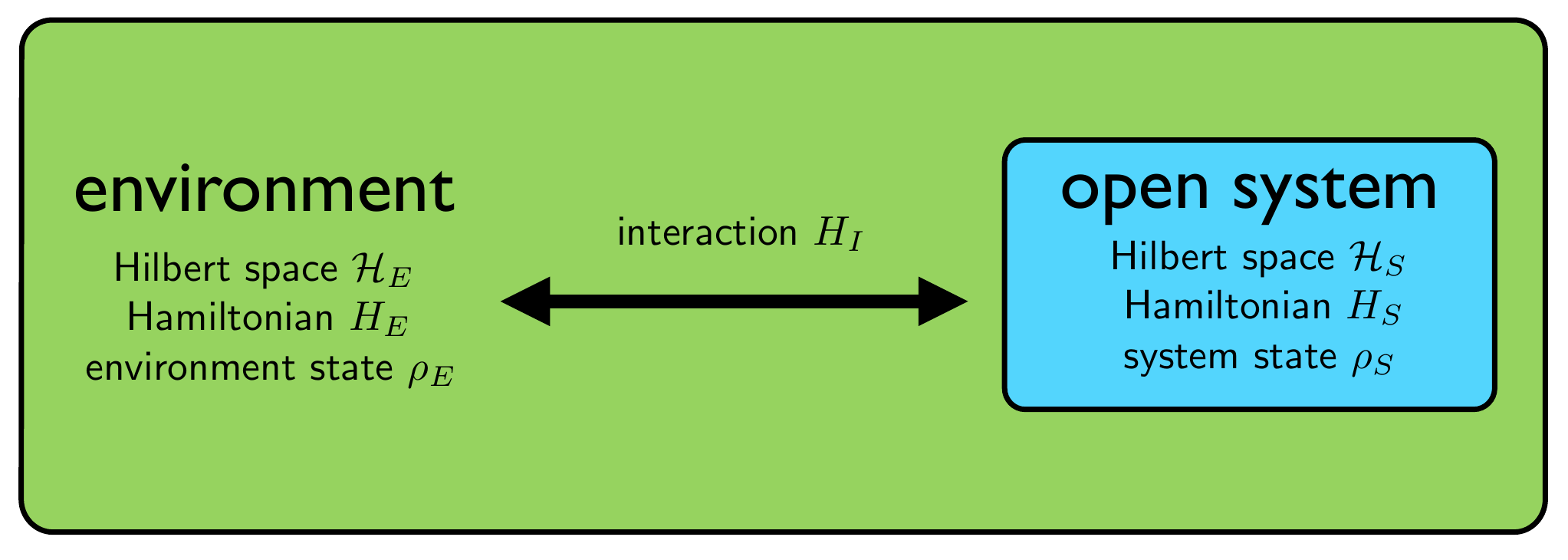}
\caption{(Color online) Sketch of an open quantum system described by
the Hilbert space ${\mathcal{H}}_S$ and the Hamiltonian $H_S$, which is
coupled to an environment with Hilbert space ${\mathcal{H}}_E$ and 
Hamiltonian $H_E$ through an interaction Hamiltonian $H_I$.}
\label{fig:open-system}
\end{figure}

An open quantum system $S$ \cite{Davies1976,Alicki1987,Breuer2002} can be 
regarded as subsystem of some larger
system composed of $S$ and another subsystem $E$, its environment, see
Fig.~\ref{fig:open-system}.
The Hilbert space of the total system $S+E$ is given by the tensor product space
\begin{equation} \label{Hilbert-total}
 {\mathcal{H}}_{SE} = {\mathcal{H}}_S\otimes{\mathcal{H}}_E,
\end{equation}
where ${\mathcal{H}}_S$ and ${\mathcal{H}}_E$ denote the Hilbert spaces of 
$S$ and $E$, respectively. Physical states of the total system are represented
by positive trace class operators $\rho_{SE}$ on ${\mathcal{H}}_{SE}$ with unit 
trace, satisfying $\rho_{SE} \geq 0$ and ${\mathrm{tr}}\rho_{SE}=1$. Given a 
state of the total system, the corresponding states of subsystems $S$ and $E$ 
are obtained by partial traces over ${\mathcal{H}}_E$ and ${\mathcal{H}}_S$, 
respectively, i.e., we have $\rho_S = {\rm{tr}}_E \rho_{SE}$ and 
$\rho_E = {\rm{tr}}_S \rho_{SE}$. We denote the convex set of 
physical states belonging to some Hilbert space ${\mathcal{H}}$ by 
${\mathcal{S}}({\mathcal{H}})$.

We suppose that the total system $S+E$ is closed and governed by a
Hamiltonian of the general form
\begin{equation} \label{Ham-total}
 H = H_S \otimes I_E + I_S \otimes H_E + H_I,
\end{equation}
where $H_S$ and $H_E$ are the free Hamiltonians of system and environment, 
respectively, and $H_I$ is an interaction Hamiltonian. The corresponding
unitary time evolution operator is thus given by
\begin{equation}
 U(t) = \exp ( -iHt ) \qquad (\hbar = 1).
\end{equation}
The dynamics of the total system states is obtained from the von Neumann 
equation,
\begin{equation} \label{Neumann}
 \frac{d}{dt}\rho_{SE}(t) = -i[H,\rho_{SE}(t)],
\end{equation}
which yields the formal solution
\begin{equation} \label{Unitary-total}
 \rho_{SE}(t) = {\mathcal{U}}_t \rho_{SE}(0) 
 \equiv U(t)\rho_{SE}(0)U^{\dagger}(t).
\end{equation}

An important concept in the theory of open quantum systems is that
of a dynamical map. To explain this concept we assume that the initial
state of the total system is an uncorrelated tensor product state
\begin{equation} \label{Product-initial-state}
 \rho_{SE}(0) = \rho_S(0) \otimes \rho_E(0),
\end{equation}
which leads to the following expression for the reduced open system state at
any time $t \geq 0$,
\begin{equation} \label{Rho-S-repr}
 \rho_S(t) = {\rm{tr}}_E \left\{ U(t) \rho_S(0) \otimes \rho_E(0) U^{\dagger}(t) 
 \right\}.
\end{equation}
Considering a fixed initial environmental state $\rho_E(0)$ and any fixed
time $t \geq 0$, Eq.~(\ref{Rho-S-repr}) defines a linear map 
\begin{equation} \label{Map-Phi}
 \Phi_t: \, S({\mathcal{H}}_S) \longrightarrow S({\mathcal{H}}_S)
\end{equation}
on the open system's state space $S({\mathcal{H}}_S)$ which maps any initial 
open system state $\rho_S(0)$ to the corresponding open system state
$\rho_S(t)$ at time $t$:
\begin{equation} \label{Dyn-map}
 \rho_S(0) \mapsto  \rho_S(t) = \Phi_t \rho_S(0).
\end{equation}
$\Phi_t$ is called quantum dynamical map. It is easy to verify that it preserves
the Hermiticity and the trace of operators, and that it is a positive map, i.e., that it
maps positive operators to positive operators. Thus, $\Phi_t$ maps physical
states to physical states.

A further important property of the dynamical map $\Phi_t$ is that it is not
only positive but also completely positive. Maps with this property are also
known as trace preserving quantum operations or quantum channels
in quantum information theory. Let us recall that a linear map 
$\Phi$ is completely positive if and only if it admits a Kraus representation
\cite{Kraus1983}, which means that there are operators $\Omega_i$ on the 
underlying Hilbert space ${\mathcal{H}}_S$ such that 
$\Phi A  = \sum_i \Omega_i A \Omega_i^{\dagger}$,
and that the condition of trace preservation takes the form 
$\sum_i \Omega_i^{\dagger}\Omega_i = I_S$.
An equivalent definition of complete positivity is the following. We consider for
any number $n=1,2,\ldots$ the tensor product space
${\mathcal{H}}_S\otimes {\mathbb C}^n$ which represents the Hilbert space
of $S$ combined with an $n$-level system. We can define a map
$\Phi\otimes I_n$ operating on the combined system by linear extension of the
relation $(\Phi\otimes I_n)(A\otimes B)= (\Phi A)\otimes B$. This map
$\Phi\otimes I_n$ thus describes a quantum operation of the composite system
which acts nontrivially only on the first factor representing subsystem $S$.
One then defines the map $\Phi$ to be $n$-positive if $\Phi\otimes I_n$ is a 
positive map, and completely positive if $\Phi\otimes I_n$ is a positive map for all 
$n$. We note that the existence of maps which are positive but not completely
positive is closely connected to the existence of entangled states. We note further
that positivity is equivalent to $1$-positivity, and that for a Hilbert space with finite 
dimension $N_S={\rm dim}\,{\mathcal{H}}_S$ complete positivity is equivalent to 
$N_S$-positivity.

If the time parameter $t$ now varies over some time interval from $0$ to $T$, 
where $T$ may be finite or infinite, we obtain a one-parameter
family of dynamical maps,
\begin{equation} \label{Family}
 \Phi = \left\{ \Phi_t \mid 0 \leq t \leq T, \Phi_0 = I \right\},
\end{equation}
where $I$ denotes the unit map, and the initial environmental state 
$\rho_E(0)$ used to construct $\Phi_t$ is still kept fixed. This family 
contains the complete information on the dynamics of all initial open system
states over the time interval $[0,T]$ we are interested in.

\subsubsection{Divisibility and time-local master equations}
\label{Div-time-local}

Let us suppose that the inverse of $\Phi_t$ exists for all times $t\geq 0$. We can 
then define a two-parameter family of maps by means of
\begin{equation} \label{two-parameter-family}
 \Phi_{t,s} = \Phi_t \Phi_s^{-1}, \qquad t\geq s \geq 0,
\end{equation}
such that we have $\Phi_{t,0}=\Phi_t$ and
\begin{equation} \label{divisibility}
 \Phi_{t,0} = \Phi_{t,s} \Phi_{s,0}.
\end{equation}
The existence of the inverse for all positive times thus allows us to
introduce the very notion of divisibility.  While $\Phi_{t,0}$ and
$\Phi_{s,0}$ are completely positive by construction, the map
$\Phi_{t,s}$ need not be completely positive and not even positive
since the inverse $\Phi^{-1}_s$ of a completely positive map $\Phi_s$
need not be positive. The family of dynamical maps is said to be
P-divisible if $\Phi_{t,s}$ is positive, and CP-divisible if
$\Phi_{t,s}$ is completely positive for all $t\geq s \geq 0$. A simple
example for a CP-divisible process is provided by a semigroup $\Phi_t
= e^{{\mathcal{L}}t}$ with a generator ${\mathcal{L}}$ in Lindblad
form. In this case we obviously have
$\Phi_{t,s}=e^{{\mathcal{L}}(t-s)}$ which is trivially completely
positive.  As we will discuss later on there are many physically
relevant models which give rise to dynamical maps which are neither
CP-divisible nor P-divisible.

An interesting property of the class of processes for which $\Phi^{-1}_t$
exists is given by the fact that they always lead to a time-local quantum
master equation for the open system states with the general structure
\begin{eqnarray} \label{tcl-master}
 \frac{d}{dt}\rho_S &=& {\mathcal{K}}_t \rho_S \\
 &=& -i\left[H_S(t),\rho_S\right] \nonumber \\
 && + \sum_{i}\gamma_{i}(t)\Big[A_{i}(t)
  \rho_S A_{i}^{\dag}(t) - \frac{1}{2}\big\{A_{i}^{\dag}(t)A_{i}(t),\rho_S\big\}\Big].
  \nonumber
\end{eqnarray}
The form of this master equation is very similar to that of a Lindblad master
equation, where however the Hamiltonian contribution $H_S(t)$, the Lindblad 
operators $A_i(t)$ (which can be supposed to be linearly independent), 
and the decay rates $\gamma_i(t)$ may depend 
on time since the process does not represent a semigroup, in general.
Note that Eq.~(\ref{tcl-master}) involves a time-dependent 
generator ${\mathcal{K}}_t$, but no convolution of the open system states
with a memory kernel as in the Nakajima-Zwanzig equation 
\cite{Nakajima1958,Zwanzig1960}. Master equations 
of the form \eqref{tcl-master} can be derived employing the
time-convolutionless projection operator technique 
\cite{Shibata1977,Shibata1979}.

It is an important open problem to formulate general necessary and sufficient 
conditions under which the master equation (\ref{tcl-master}) leads to a 
completely positive dynamics. If the process represents a semigroup, the
Hamiltonian $H_S$, the operators $A_i$ and the rates $\gamma_i$ must be 
time-independent and a necessary and
sufficient condition for complete positivity of the dynamics is simply that all
decay rates are positive, $\gamma_i\geq 0$. This is the famous
Gorini-Kossakowski-Sudarshan-Lindblad theorem 
\cite{Gorini1976a,Lindblad1976a}. However, we will see later
by several examples that in the time-dependent case the rates 
$\gamma_i(t)$ can indeed become temporarily negative without violating
the complete positivity of the dynamics.

On the other hand, for divisible quantum processes it is indeed possible to 
formulate necessary and sufficient conditions. In fact, the master equation 
(\ref{tcl-master}) leads to a CP-divisible dynamics if and only if all rates are 
positive for all times, $\gamma_i(t)\geq 0$, which follows from a straightforward
extension of the Gorini-Kossakowski-Sudarshan-Lindblad theorem. Moreover,
the master equation (\ref{tcl-master}) leads to a P-divisible dynamics if and only if 
the weaker conditions
\begin{equation} \label{P-divisibilty-condition}
 \sum_i \gamma_i(t) |\langle n|A_i(t)|m\rangle |^2 \geq 0
\end{equation}
hold for all orthonormal bases $\{ |n\rangle\}$ of the open system and all
$n\neq m$. This statement can be obtained by applying a characterization
of the generators of positive semigroups due to 
\textcite{Kossakowski1972a,Kossakowski1972b}.

\subsection{Classical versus quantum non-Markovianity}

\subsubsection{Classical stochastic processes and the Markov condition}
\label{stoch-proc-Markov-cond}

In classical probability theory \cite{Gardiner1985,Kampen1992} a stochastic 
process $X(t)$, $t\geq 0$, taking values in a
discrete set $\{x_i\}_{i\in{\mathbb N}}$ is characterized by a hierarchy of joint 
probability distributions $P_n=P_n(x_n,t_n;x_{n-1},t_{n-1};\ldots;x_1,t_1)$ for all 
$n\in{\mathbb N}$ and times $t_n\geq t_{n-1}\geq\ldots\geq t_1\geq 0$.
The distribution $P_n$ yields
the probability that the process takes on the value $x_1$ at time $t_1$, the
value $x_2$ at time $t_2$, \ldots, and the value $x_n$ at time $t_n$. In order for
such a hierarchy to represent a stochastic process the Kolmogorov consistency
conditions must be satisfied which, apart from conditions of positivity and 
normalization, require in particular the relation
\begin{eqnarray} \label{Kolmogorov-consistency}
 \lefteqn{
 \sum_{x_m} P_n(x_n,t_n;\ldots;x_m,t_m;\ldots;x_1,t_1) } \nonumber \\
 && = P_{n-1}(x_n,t_n;\ldots;x_1,t_1),
\end{eqnarray}
connecting the $n$-point probability distribution $P_n$ to the $(n-1)$-point 
probability distribution $P_{n-1}$. 

A stochastic process $X(t)$ is said to be Markovian if the conditional probabilities
defined by
\begin{eqnarray} \label{conditional-probabilities}
 \lefteqn{
 P_{1|n}(x_{n+1},t_{n+1}|x_n,t_n;\ldots;x_1,t_1) } \nonumber \\
 && = \frac{P_{n+1}(x_{n+1},t_{n+1};\ldots;x_1,t_1)}{P_n(x_n,t_n;\ldots;x_1,t_1)}
\end{eqnarray}
satisfy the relation
\begin{eqnarray} \label{classical-Markov-condition}
 \lefteqn{
 P_{1|n}(x_{n+1},t_{n+1}|x_n,t_n;\ldots;x_1,t_1) } \nonumber \\
 && = P_{1|1}(x_{n+1},t_{n+1}|x_n,t_n).
\end{eqnarray}
This is the classical Markov condition which means that the probability for the
stochastic process to take the value $x_{n+1}$ at time $t_{n+1}$, under the 
condition that it assumed values $x_i$ at previous times $t_i$, only depends on
the last previous value $x_n$ at time $t_n$. In this sense the process
is said to have no memory, since the past history prior to $t_n$ is irrelevant
to determine the future once we know the value $x_n$ the process assumed
at time $t_n$. Note that Eq.~(\ref{classical-Markov-condition}) imposes an
infinite number of conditions for all $n$-point probability distributions which cannot
be checked if only a few low-order distributions are known.

The Markov condition (\ref{classical-Markov-condition}) substantially simplifies the
mathematical description of stochastic processes. In fact, one can show that
under this condition the whole hierarchy of joint probability distributions can be
reconstructed from the initial $1$-point distribution $P_1(x_0,0)$ and the
conditional transition probability 
\begin{equation} \label{def-T}
 T(x,t|y,s) \equiv P_{1|1}(x,t|y,s)
\end{equation}
by means of the relations
\begin{eqnarray} \label{classical-Markov-process}
 \lefteqn{
 P_n(x_n,t_n;x_{n-1},t_{n-1};\ldots;x_1,t_1) } \nonumber \\
 && = \prod_{i=1}^{n-1}  T(x_{i+1},t_{i+1}|x_{i},t_{i}) P_1(x_1,t_1)
\end{eqnarray}
and
\begin{equation} \label{1-point-distr}
 P_1(x_1,t_1) = \sum_{x_0} T(x_1,t_1|x_0,0) P_1(x_0,0).
\end{equation}
For a Markov process the transition probability has to obey the
Chapman-Kolmogorov equation
\begin{equation} \label{Chapman-Kolmogorov}
 T(x,t|y,s) = \sum_z T(x,t|z,\tau) T(z,\tau|y,s)
\end{equation}
for $t\geq\tau\geq s$. Thus, a classical Markov process is uniquely characterized 
by a probability distribution for the initial states of the process and a conditional 
transition probability satisfying the Chapman-Kolmogorov equation
(\ref{Chapman-Kolmogorov}). Indeed, the latter
provides the necessary condition in order to ensure that the joint
probabilities defined by \eqref{classical-Markov-process} satisfy the
condition \eqref{Kolmogorov-consistency}, so that they actually define a classical 
Markov process. Provided the conditional transition 
probability is differentiable with respect to time (which will always be assumed 
here) one obtains an equivalent differential equation, namely the 
Chapman-Kolmogorov equation:
\begin{eqnarray} \label{diff-Chapman-Kolmogorov}
 \lefteqn{
 \frac{d}{dt} T(x,t|y,s) } \\
 && = \sum_z \Big[ W_{xz}(t) T(z,t|y,s) - W_{zx}(t) T(x,t|y,s) \Big],
 \nonumber
\end{eqnarray}
where $W_{zx}(t)\geq 0$, representing the rate (probability per unit of time) for a
transition to the state $z$ given that the state is $x$ at time $t$. An equation of
the same structure holds for the $1$-point probability distribution of the process:
\begin{equation} \label{classical-Pauli-master-equation}
 \frac{d}{dt} P_1(x,t)
 = \sum_z \Big[ W_{xz}(t) P_1(z,t) - W_{zx}(t) P_1(x,t) \Big],
\end{equation}
which is known as Pauli master equation for a classical Markov
process. The conditional transition probability of the process can be
obtained solving Eq.~\eqref{classical-Pauli-master-equation}
with the initial condition $P_1(x,s)=\delta_{xy}$.

\subsubsection{Non-Markovianity in the quantum regime}
\label{NM-quantum-regime}

The above definition of a classical Markov process cannot be transferred 
immediately to the quantum regime \cite{Vacchini2011a}. In order to illustrate the 
arising difficulties let 
us consider an open quantum system as described in Sec.~\ref{Basic-Concepts}. 
Suppose we carry out projective measurements of an open system observable 
$\hat{X}$ at times $t_n\geq t_{n-1}\geq\ldots\geq t_1\geq 0$. For simplicity 
we assume that this observable is non-degenerate with spectral decomposition
$\hat{X}=\sum_x x |\varphi_x\rangle\langle\varphi_x|$. As in 
Eq.~(\ref{Unitary-total}) we write the
unitary evolution superoperator as 
${\mathcal{U}}_t\rho_{SE}=U_t\rho_{SE} U_t^{\dagger}$
and the quantum operation corresponding to the measurement outcome $x$ as 
${\mathcal{M}}_x\rho_{SE}
= |\varphi_x\rangle\langle\varphi_x|\rho_{SE}|\varphi_x \rangle\langle\varphi_x|$. 
Applying the Born rule and the projection postulate one 
can then write a joint probability distribution
\begin{eqnarray} \label{qm-joint-distr}
 \lefteqn{
 P_n(x_n,t_n;x_{n-1},t_{n-1};\ldots;x_1,t_1) } \nonumber \\
 && = {\mathrm{tr}} \left\{
 {\mathcal{M}}_{x_n} {\mathcal{U}}_{t_n-t_{n-1}} 
 \ldots {\mathcal{M}}_{x_1} {\mathcal{U}}_{t_1} 
 \rho_{SE}(0) \right\}
\end{eqnarray}
which yields the probability of observing a certain sequence
$x_n,x_{n-1},\ldots,x_1$ of measurement outcomes at the respective
times $t_n,t_{n-1},\ldots,t_1$. Thus, it is of course possible in
quantum mechanics to define a joint probability distribution for any
sequence of measurement results. However, as is well known these
distribution do in general not satisfy condition
(\ref{Kolmogorov-consistency}) since measurements carried out at
intermediate times in general destroy quantum interferences. 
The fact that the joint probability distributions (\ref{qm-joint-distr}) violate
in general the Kolmogorov condition (\ref{Kolmogorov-consistency}) is even
true for closed quantum system. 
Thus, the quantum joint probability distributions given by
(\ref{qm-joint-distr}) do not represent a classical hierarchy of joint
probabilities satisfying the Kolmogorov consistency conditions.
More generally, one can consider other joint probability
distributions corresponding to different quantum operations
describing non-projective, generalized measurements.

For an open system coupled to some 
environment measurements performed on the open system not only influence 
quantum interferences but also system-environment correlations. For example, 
if the system-environment state prior to a projective measurement at time $t_i$ 
is given by $\rho_{SE}(t_i)$, the state after the measurement 
conditioned on the outcome $x$ is given by
\begin{equation} \label{state-reduction}
 \rho'_{SE}(t_i) = 
 \frac{{\mathcal{M}}_{x}\rho_{SE}(t_i)}
 {{\mathrm{tr}}{\mathcal{M}}_{x}\rho_{SE}(t_i)}
 = |\varphi_x\rangle\langle\varphi_x| \otimes \rho^x_{E}(t_i),
\end{equation}
where $\rho^x_{E}$ is an environmental state which may depend on the
measurement result $x$. Hence, projective measurements completely
destroy system-environment correlations, leading to an uncorrelated
tensor product state of the total system, and, therefore, strongly influence the
subsequent dynamics.

We conclude that an {\textit{intrinsic}} characterization and quantification 
of memory effects in the dynamics of open quantum systems, 
which is independent of any prescribed measurement scheme influencing 
the time evolution, has to be based solely on the properties of the dynamics of
the open system's density matrix $\rho_S(t)$.

\subsection{Quantum non-Markovianity and information flow}
\label{q-NM-info-flow}

The first approach to quantum non-Markovianity to be discussed here
is based on the idea that memory effects in the dynamics of open systems
are linked to the exchange of information between the open system and its
environment: While in a Markovian process the open system continuously looses
information to the environment, a non-Markovian process is characterized by
a flow of information from the environment back into the open system 
\cite{Breuer2009b,Laine2010a}. 
In such a way quantum non-Markovianity is associated to a notion of
quantum memory, namely information which has been transferred to the
environment, in the form of system-environment correlations or changes in the
environmental states, and is later retrieved by the system.
To make this idea more precise we employ an appropriate distance measure 
for quantum states.

\subsubsection{Trace distance and distinguishability of quantum states}
\label{Trace-distance-distinguishability}

The trace norm of a trace class operator $A$ is defined by 
$||A||={\mathrm{tr}}|A|$,
where the modulus of the operator is given by $|A|=\sqrt{A^{\dagger}A}$.
If $A$ is selfadjoint the trace norm can be expressed as the sum of the
moduli of the eigenvalues $a_i$ of $A$ counting multiplicities, $||A||=\sum_i |a_i|$.
This norm leads to a natural measure for the distance between
two quantum states $\rho^1$ and $\rho^2$ known as trace distance:
\begin{equation}\label{trace-distance}
 D(\rho^1,\rho^2) = \frac{1}{2} ||\rho^1-\rho^2||.
\end{equation}
The trace distance represents a metric on the state space 
${\mathcal{S}}({\mathcal{H}})$ of the underlying Hilbert space ${\mathcal{H}}$. 
We have the bounds
$0\leq D(\rho^1,\rho^2)\leq 1$, where $D(\rho^1,\rho^2)=0$ if and only if 
$\rho^1=\rho^2$, and $D(\rho^1,\rho^2)=1$ if and only if $\rho^1$ and $\rho^2$ 
are orthogonal. Recall that two density matrices are said to be orthogonal if their 
supports, i.e. the subspaces spanned by their eigenstates with nonzero 
eigenvalue, are orthogonal. The trace distance has a series of interesting 
mathematical and physical features which make it a useful distance measure in 
quantum information theory \cite{Nielsen2000}. 
There are two properties which are the most relevant for our purposes.

The first property is that the trace distance between two quantum states admits
a clear physical interpretation in terms of the distinguishability  of these states.
To explain this interpretation consider two parties, Alice and Bob, and suppose 
that Alice prepares a quantum system in one of two states, $\rho^1$ or $\rho^2$, 
with a probability of $\frac{1}{2}$ each, and sends the system to Bob. The task of
Bob is to find out by means of a single quantum measurement whether the 
system is in the state $\rho^1$ or $\rho^2$. It can be shown that the maximal 
success probability Bob can achieve through an optimal strategy is
directly connected to the trace distance:
\begin{equation} \label{P-max}
 P_{\max} = \frac{1}{2} \left[1+ D(\rho^1,\rho^2)\right].
\end{equation}
Thus, we see that the trace distance represents the bias in
favor of a correct state discrimination by Bob. For this reason the trace 
distance $D(\rho^1,\rho^2)$ can be interpreted as the distinguishability of the 
quantum states $\rho^1$ and $\rho^2$. For example, suppose the states 
prepared by Alice are orthogonal such that we have $D(\rho^1,\rho^2)=1$. In this 
case we get $P_{\max}=1$, which is a well known fact since orthogonal states 
can be distinguished with certainty by a single measurement, an optimal strategy
of Bob being to measure the projection onto the support of $\rho^1$ or $\rho^2$.

\begin{figure}[tbh]
\includegraphics[width=0.45\textwidth]{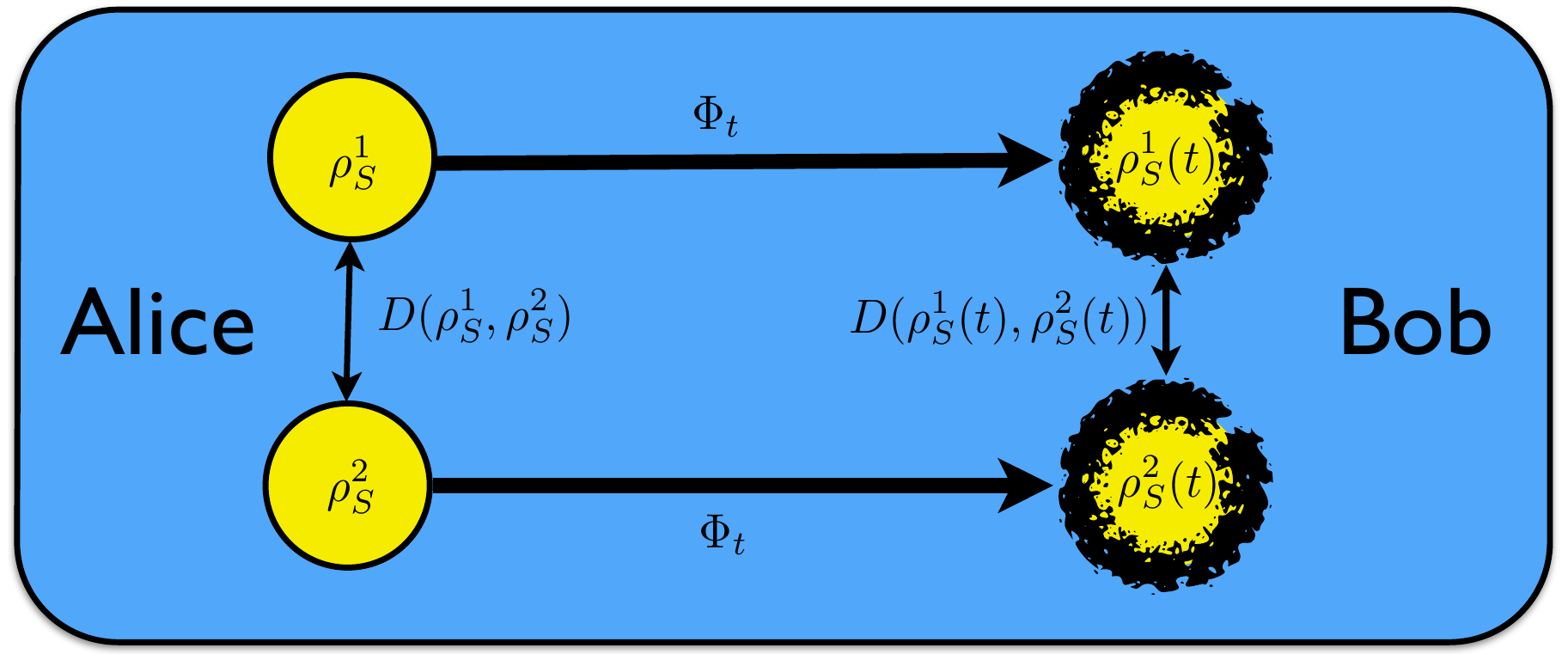}
\caption{(Color online) Illustration of the
loss of distinguishability quantified by the trace distance $D$ between
density matrices as a consequence of the action of a quantum
dynamical map $\Phi_t$. Alice prepares two distinct states, but a dynamical map
acts as a noisy channel, thus generally reducing the information
available to Bob in order to distinguish among the states by performing  
measurements on the system only.}
\label{fig:info-loss}
\end{figure}

The second important property of the trace distance is given by the fact that
any completely positive and trace preserving map $\Lambda$ is a contraction
for the trace distance, i.e. we have
\begin{equation} \label{contraction-property}
 D(\Lambda\rho^1,\Lambda\rho^2) \leq D(\rho^1,\rho^2)
\end{equation}
for all states $\rho^{1,2}$. In view of the interpretation of the trace distance 
we thus conclude that a trace preserving quantum operation can never increase 
the distinguishability of any two quantum states. We remark that the equality sign
in (\ref{contraction-property}) holds if $\Lambda$ is a unitary 
transformation, and that (\ref{contraction-property}) is also valid for trace 
preserving maps which are positive but not completely positive.

\subsubsection{Definition and quantification of memory effects}
\label{def-NM-measure}

In the preceding subsection we have interpreted the trace distance of two
quantum states as the distinguishability of these states, where it has been
assumed that the quantum state Bob receives is identical to the state
prepared by Alice. Suppose now that Alice prepares her states $\rho_S^{1,2}(0)$ 
as initial states of an open quantum system $S$ coupled to some 
environment $E$. Bob will then receive at time $t$ the system in one of the states 
$\rho_S^{1,2}(t)=\Phi_t\rho_S^{1,2}(0)$, where $\Phi_t$ denotes the 
corresponding quantum dynamical map. This construction is equivalent to Alice 
sending her states through a noisy quantum channel described by the completely 
positive and trace preserving map $\Phi_t$. According to 
(\ref{contraction-property}) the dynamics generally diminishes the trace distance 
and, therefore, the distinguishability of the states,
\begin{equation} \label{decrease}
 D(\rho_S^1(t),\rho_S^2(t)) \leq D(\rho_S^1(0),\rho_S^2(0)),
\end{equation}
such that it will in general be harder for Bob to discriminate the states prepared
by Alice.  This fact is illustrated schematically in
Fig.~\ref{fig:info-loss}, in which the distinguishability of states
prepared by Alice decreases due to the action of a dynamical map, so
that Bob is less able to discriminate among the two. Thus, we can interpret any 
decrease of the trace distance 
$D(\rho_S^1(t),\rho_S^2(t))$ as a loss of information from the open system into 
the environment. Conversely, if the trace distance $D(\rho_S^1(t),\rho_S^2(t))$ 
increases, we say that information flows back from the environment into the open 
system.

This interpretation naturally leads to the following definition: A quantum process
given by a family of quantum dynamical maps $\Phi_t$ is said to be
Markovian if the trace distance $D(\rho_S^1(t),\rho_S^2(t))$ corresponding to all
pair of initial states $\rho_S^1(0)$ and $\rho_S^2(0)$ decreases monotonically for
all times $t\geq 0$. Quantum Markovian behavior thus means a continuous
loss of information from the open system to the environment. Conversely, a
quantum process is non-Markovian if there is an initial pair of states
$\rho_S^1(0)$ and $\rho_S^2(0)$ such that the trace distance 
$D(\rho_S^1(t),\rho_S^2(t))$
is non-monotonic, i.e. starts to increase for some time $t>0$. If this happens,
information flows from the environment back to the open system, which clearly
expresses the presence of memory effects: Information contained in the open
system is temporarily stored in the environment and comes back at a later time
to influence the system.
A crucial feature of this definition of quantum Markovian process is the fact
that non-Markovianity can be directly experimentally assessed, provided
one is able to perform tomographic measurement of different initial
states at different times during the evolution. No prior information
on the dynamical map $\Phi_t$ is required, apart from its very
existence.

\begin{figure}[tbh]
\includegraphics[width=0.45\textwidth]{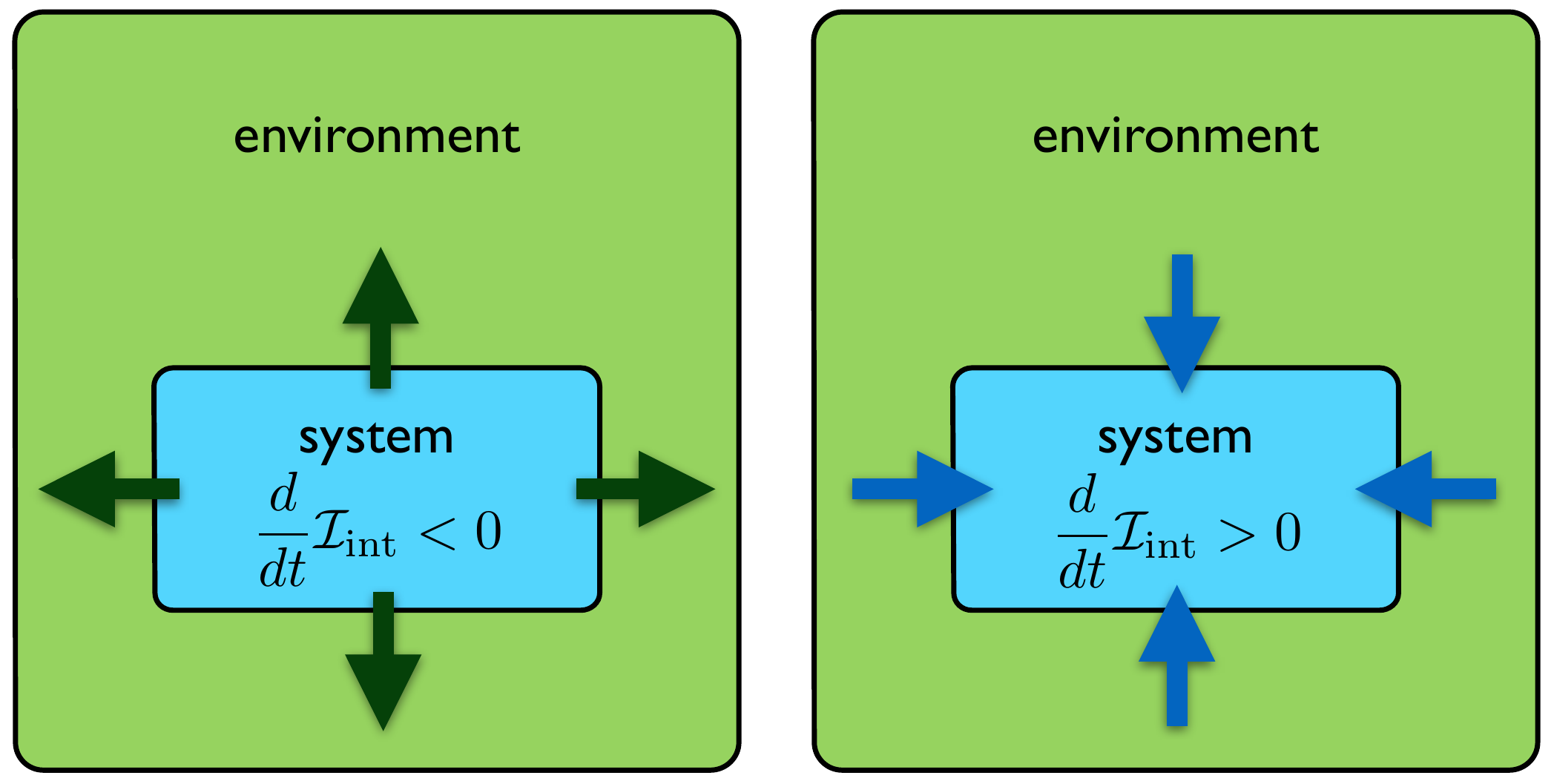}
\caption{(Color online) Illustration of the information flow between an open system 
and its environment according to Eq.~(\ref{relation-info-flow}). Left: The open 
system looses information to the environment, corresponding to a decrease of 
$\mathcal{I}_{\rm int}(t)$ and Markovian dynamics. 
Right: Non-Markovian dynamics is characterized by a backflow of information
from the environment to the system and a corresponding increase 
of $\mathcal{I}_{\rm int}(t)$.}
\label{fig:info-flow}
\end{figure}

In order to explain in more detail this interpretation in terms of an information
flow between system and environment let us define the quantities
\begin{align}
  \label{eq:1}
  \mathcal{I}_{\rm int}(t)&=D(\rho_S^1(t),\rho_S^2(t))\\
 \intertext{and}
  \label{eq:1bis}
 \mathcal{I}_{\rm ext}(t)&=
 D(\rho_{SE}^1(t),\rho_{SE}^2(t)) - D(\rho_S^1(t),\rho_S^2(t)).
\end{align} 
Here, $\mathcal{I}_{\rm int}(t)$ is the distinguishability of the open system states 
at time $t$, while $\mathcal{I}_{\rm ext}(t)$ is the distinguishability of the total 
system states minus the distinguishability of the open system states at time $t$. In 
other words, $\mathcal{I}_{\rm ext}(t)$ can be viewed as the gain in the state 
discrimination Bob could
achieve if he were able to carry out measurements on the total system instead
of measurements on the open system only.
We can therefore interpret $\mathcal{I}_{\rm int}(t)$ as the information inside the 
open system and $\mathcal{I}_{\rm ext}(t)$ as the amount of information which 
lies outside the open system, i.e., as the information which is not accessible when 
only measurements on the open system can be performed. Obviously, we have
$\mathcal{I}_{\rm int}(t)\geq 0$ and $\mathcal{I}_{\rm ext}(t)\geq 0$. Since the 
total system dynamics is unitary the distinguishability of the total system states is
constant in time. Moreover we have 
$D(\rho_{SE}^1(0),\rho_{SE}^2(0))=D(\rho_S^1(0),\rho_S^2(0))$ because, by
assumption, the initial total states are uncorrelated with the same reduced 
environmental state. Hence, we obtain
\begin{equation}\label{relation-info-flow}
 \mathcal{I}_{\rm int}(t) + \mathcal{I}_{\rm ext}(t) 
 = \mathcal{I}_{\rm int}(0) = {\rm const.}
\end{equation}
Thus, initially there is no information outside the open system, 
$\mathcal{I}_{\rm ext}(0)=0$. If $\mathcal{I}_{\rm int}(t)$ decreases 
$\mathcal{I}_{\rm ext}(t)$ must increase and vice versa, 
which clearly expresses the idea of the exchange of information between 
the open system and the environment illustrated in Fig.~\ref{fig:info-flow}. 
Employing the properties of the trace distance one can derive the following 
general inequality \cite{Laine2010b} which holds for all $t\geq 0$:
\begin{eqnarray} \label{bound-I-out}
 \mathcal{I}_{\rm ext}(t)
 && \leq D(\rho_{SE}^1(t),\rho_S^1(t)\otimes\rho_E^1(t)) \\
 && \quad + D(\rho_{SE}^2(t),\rho_S^2(t)\otimes\rho_E^2(t))
 + D(\rho_E^1(t),\rho_E^2(t)). \nonumber
\end{eqnarray}
The right-hand side of this inequality consists of three terms: The first two terms 
provide a measure for the correlations in the total system states 
$\rho_{SE}^{1,2}(t)$, given by the trace distance between these states and the 
product of their marginals, while the third term is the trace distance between the 
corresponding environmental states. Thus, when $\mathcal{I}_{\rm ext}(t)$
increases over the initial value $\mathcal{I}_{\rm ext}(0)=0$ system-environment 
correlations are built up or the environmental states become different, implying an 
increase of the distinguishability of the environmental states. This clearly 
demonstrates that the corresponding decrease of the distinguishability 
$\mathcal{I}_{\rm int}(t)$ of the open system states always has an impact on 
degrees of freedom which are inaccessible by measurements on the open 
system. Conversely, if $\mathcal{I}_{\rm int}(t)$ starts to increase at some point of 
time $t$, the corresponding decrease of $\mathcal{I}_{\rm ext}(t)$
implies that system-environment correlations must be present already at time $t$, 
or that the environmental states are different at this point of time.

On the ground of the above definition for non-Markovian dynamics one
is naturally led to introduce the following measure for the degree of memory 
effects
\begin{equation} \label{NM-measure}
 \mathcal{N}(\Phi) = \max_{\rho^{1,2}_S} 
 \int_{\sigma>0} dt \; \sigma(t),
\end{equation}
where
\begin{equation} \label{trace-distance-deriv}
\sigma(t) \equiv \frac{d}{dt} D\left(\Phi_t\rho^1_S,\Phi_t\rho^2_S\right)
\end{equation}
denotes the time derivative of the trace distance of the evolved
pair of states. In Eq.~(\ref{NM-measure}) the time integral is extended over
all intervals in which $\sigma(t) > 0$, i.e., in which the trace distance increases
with time, and the maximum is taken over all pairs of initial states 
$\rho^{1,2}_S$ of the open system's state space 
${\mathcal{S}}({\mathcal{H}}_S)$.
Thus, $\mathcal{N}(\Phi)$ is a positive functional of the family of dynamical maps 
$\Phi$ which represents a measure for the maximal total flow of information
from the environment back to the open system. By construction we have
$\mathcal{N}(\Phi)=0$ if and only if the process is Markovian.

The maximization over all pairs of quantum states in Eq.~(\ref{NM-measure})
can be simplified substantially employing several important properties of
the functional $\mathcal{N}(\Phi)$ and the convex structure of the set of quantum 
states. A certain pair of states $\rho^{1,2}_S$
is said to be an optimal state pair if the maximum in Eq.~(\ref{NM-measure})
is attained for this pair of states. Thus, optimal state pairs lead to the maximal
possible backflow of information during their time evolution. One can show that
optimal states pairs $\rho^{1,2}_S$ lie on the boundary of the state
space, and are in particular always orthogonal \cite{Wissmann2012a}. This is a 
quite natural result in view of the interpretation in terms of an information flow 
since orthogonality implies that $D(\rho^1_S,\rho^2_S)=1$, which shows that 
optimal state pairs have maximal initial distinguishability, corresponding to a 
maximal amount of initial information. 
An even more drastic simplification is obtained by employing the following
equivalent representation \cite{Liu2014a}:
\begin{equation} \label{local-repr}
 \mathcal{N}(\Phi) = \max_{\rho\in\partial U(\rho_0)} 
 \int_{\bar{\sigma}>0} dt ~\bar{\sigma}(t),
\end{equation}
where
\begin{equation} \label{scaled-trace-distance-deriv}
\bar{\sigma}(t) \equiv
 \frac{\frac{d}{dt}D\left(\Phi_t\rho,\Phi_t\rho_0\right)}{D\left(\rho,\rho_0\right)}
\end{equation}
is the time derivative of the trace distance at time $t$ divided by the initial
trace distance. In Eq.~(\ref{local-repr}) $\rho_0$ is a fixed point of the 
interior of the state space and $\partial U(\rho_0)$ an arbitrary surface
in the state space enclosing, but not containing $\rho_0$. The maximization is 
then taken over all points of such an enclosing surface. This representation
often significantly simplifies the analytical, numerical or experimental 
determination of the measure since it only involves a maximization over a single 
input state $\rho$. It is particularly advantageous if the open system dynamics has 
an invariant state and if this state is taken to be $\rho_0$, such that only one state 
of the pair evolves in time. Equation (\ref{local-repr}) may be called local 
representation for it shows that optimal state pairs can be found in any local 
neighborhood of any interior point of the state space. Furthermore, the representation 
reveals the universality of memory effects, namely that for a non-Markovian 
dynamics memory effects can be observed everywhere in state space.

\subsubsection{Generalizing the trace distance measure}
\label{gen-trace-distance-measure}

There is an interesting generalization of the above definition and quantification of 
non-Markovian quantum dynamics first suggested by \textcite{Chruscinski2011a}. 
Returning to the interpretation of the
trace distance described in Sec.~\ref{Trace-distance-distinguishability}, we
may suppose that Alice prepares her quantum system in the states $\rho_S^1$ 
or $\rho_S^2$ with corresponding probabilities $p_1$ and $p_2$,
where $p_1+p_2=1$. Thus, we assume that Alice gives certain
weights to her states which need not be equal. In this case one can again 
derive an expression for the maximal probability for Bob to identify the state 
prepared by Alice:
\begin{equation} \label{P-max-generalized}
 P_{\max} = \frac{1}{2} \left(1+ \|p_1\rho_S^1-p_2\rho_S^2\|\right).
\end{equation}
As follows from \eqref{P-max-generalized} the quantity 
$\|p_1\rho_S^1-p_2\rho_S^2\|$ gives the bias in favor of
the correct identification of the state prepared by Alice. We note that the operator
$\Delta =p_1\rho_S^1-p_2\rho_S^2$ is known as Helstrom matrix 
\cite{Helstrom1967a}, and
that Eq.~(\ref{P-max-generalized}) reduces to Eq.~(\ref{P-max}) in the
unbiased case $p_1=p_2=\frac{1}{2}$. Note that the interpretation 
in terms of an information flow between system and environment still holds for
the biased trace distance, as can be seen replacing in \eqref{eq:1}  and
\eqref{eq:1bis} the trace distance by the norm of the Helstrom matrix. 

Following this approach we can define a process to be Markovian if the 
function
$\| \Phi_t(p_1\rho^1_S-p_2\rho^2_S)\|$ 
decreases
monotonically with time for all $p_{1,2}$ and all pairs of initial open system states 
$\rho^{1,2}_S$. Consequently, the extended measure for non-Markovianity is 
then given by
\begin{equation} \label{NM-extended-measure}
 \mathcal{N}(\Phi) = \max_{p_i,\rho^i_S} 
 \int_{\sigma>0} dt \; \sigma(t),
\end{equation}
where
\begin{equation} \label{sigma-extended-measure}
 \sigma(t) \equiv \frac{d}{dt} \|\Phi_t(p_1\rho^1_S-p_2\rho^2_S)\|.
\end{equation}
This generalized definition of quantum non-Markovianity leads to several 
important conclusions. First, we note that also for the measure
(\ref{NM-extended-measure}) optimal state pairs are orthogonal,
satisfying $\|p_1\rho_S^1-p_2\rho_S^2\|=1$ and, therefore, corresponding to
maximal initial distinguishability. 
The maximum in (\ref{NM-extended-measure}) can therefore be taken over all 
Hermitian (Helstrom) matrices $\Delta$ with unit
trace norm. This leads to a local representation analogous to 
Eq.~(\ref{local-repr}):
\begin{equation}
 \mathcal{N}(\Phi) = \max_{\Delta\in\partial U(0)} 
 \int_{\bar{\sigma}>0} dt \; \bar{\sigma}(t),
\end{equation}
where $\bar{\sigma}(t) =  \frac{d}{dt} \|\Phi_t \Delta \| / \|\Delta\|$,
and $\partial U(0)$ is a closed surface which encloses the point $\Delta=0$ in the
${\mathbb R}$-linear vector space of Hermitian matrices.

\begin{figure}[tbh]
\includegraphics[width=0.38\textwidth]{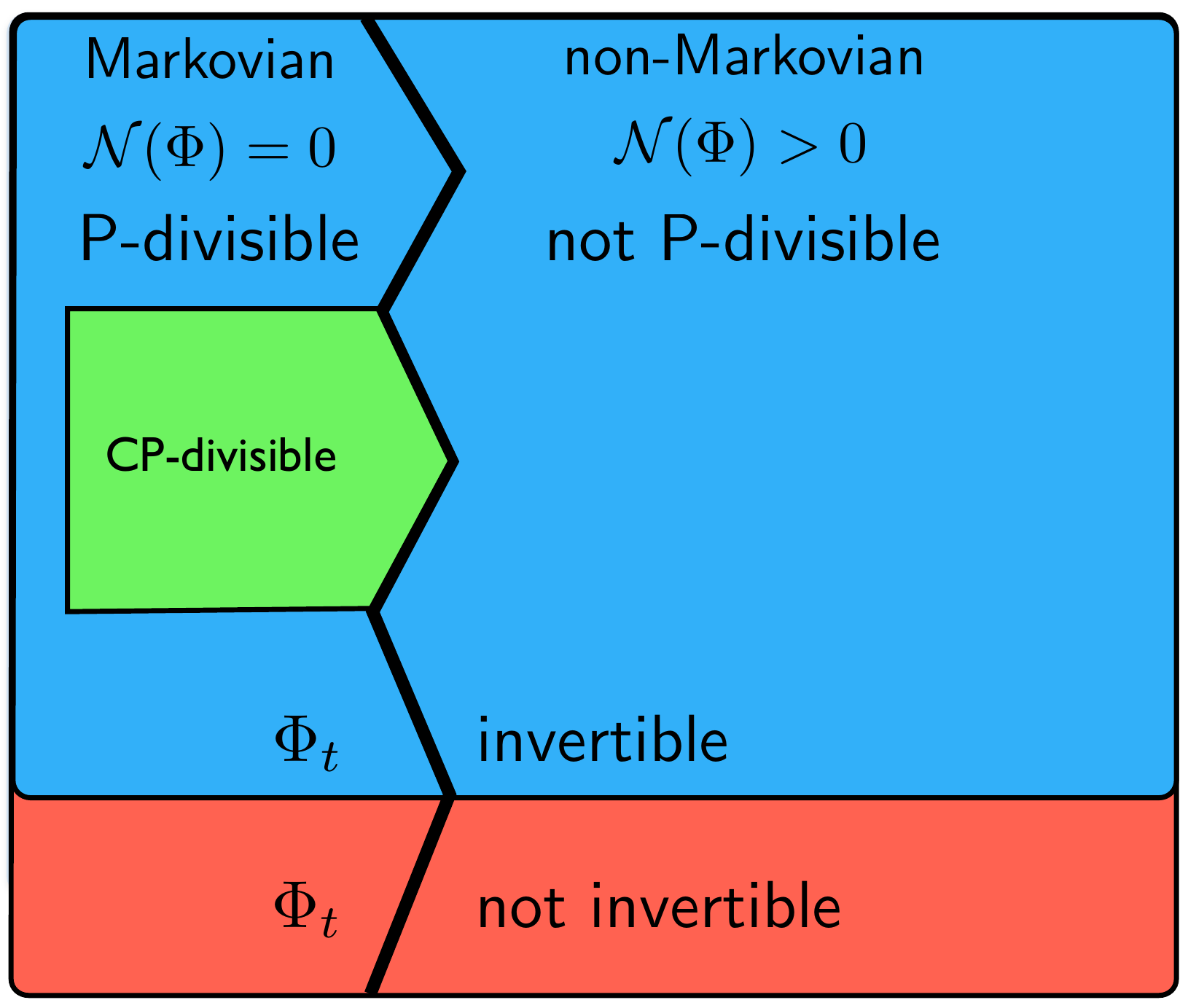}
\caption{(Color online) Schematic picture of the relations between the
concepts of quantum Markovianity and divisibility of the family of quantum 
dynamical maps $\Phi_t$.  The blue area to the left of the thick black line 
represents the set of Markovian quantum processes for which the inverse of the 
dynamical map exists, which is identical to the set of P-divisible processes.  This 
set contains as subset the set of CP-divisible processes, depicted in green. The 
blue area to the right represents the non-Markovian processes which are 
therefore neither CP nor P-divisible. The zig-zag line dividing the two regions 
points to the fact that neither of them is convex. The red areas mark 
the Markovian or non-Markovian processes for which the inverse of $\Phi_t$ does 
not exist.}
\label{fig:scheme}
\end{figure}

Another important conclusion is obtained if we assume, as was done in
Sec.~\ref{Div-time-local}, that $\Phi_t^{-1}$ exists. It can be shown
that under this condition the quantum process is Markovian if and only
if $\Phi_t$ is P-divisible, which follows immediately from a theorem
by \textcite{Kossakowski1972a,Kossakowski1972b}.
The relationships between Markovianity and divisibility of the
dynamical map are illustrated in Fig.~\ref{fig:scheme},
where also the situation in which the inverse of $\Phi_t$ as a linear map does not
exist for all times has been considered.
The generalized characterization of non-Markovianity is
sensitive to features of the open system dynamics neglected by the
trace distance measure \cite{Wang2013a}.  It is worth stressing that
in keeping with the spirit of the approach, this generalized measure
of non-Markovianity, in addition to its clear connection with the
mathematical property of P-divisibility of the dynamical map, can
still be directly evaluated by means of experiments. More specifically,
both the quantification of non-Markovianity according to the
measure \eqref{NM-measure}, as well as its generalization
\eqref{NM-extended-measure} can be obtained from the very same data.

\subsubsection{Connection between quantum and classical non-Markovianity}
\label{quantum-classical-connection}

The definition of quantum non-Markovianity of
Sec.~\ref{gen-trace-distance-measure} allows to establish an immediate
general connection to the classical definition. Suppose as in
Sec.~\ref{Div-time-local} that the inverse of the dynamical maps
$\Phi_t$ exists such that we have the time-local quantum master
equation (\ref{tcl-master}), in which for the sake of simplicity we
consider only the rates to be time dependent. 
Suppose further that $\Phi_t$ maps operators which are diagonal in a fixed
basis $\{|n\rangle\}$ to operators diagonal in the same basis.
If $\rho_S(0)$ is an initial state diagonal in this basis, then $\rho_S(t)$ can be 
expressed at any time in the form:
\begin{equation}\label{rhoeigen}
 \rho_S(t) = \sum_n P_n(t) |n\rangle\langle n|,
\end{equation}
where $P_n(t)$ denote the time-dependent eigenvalues.  The quantum
master equation then leads to the following equation of motion for the
probabilities $P_n(t)$:
\begin{equation} \label{quantum-pauli-meq}
 \frac{d}{dt} P_n(t) = \sum_m \Big[ W_{nm}(t) P_m(t) - W_{mn}(t) P_n(t) \Big],
\end{equation}
where
\begin{equation}\label{quantum-pauli-rates}
 W_{nm}(t) 
 = \sum_i \gamma_i(t) |\langle n|A_i|m\rangle |^2.
\end{equation}
We observe that Eq.~(\ref{quantum-pauli-meq}) has exactly the structure of a
classical Pauli master equation (\ref{classical-Pauli-master-equation}).
Therefore, Eq.~(\ref{quantum-pauli-meq}) can be interpreted as differential 
Chapman-Kolmogorov equation (\ref{diff-Chapman-Kolmogorov}) for the 
conditional transition probability of a classical Markov process if and only if the 
rates (\ref{quantum-pauli-rates}) are positive. As we have seen a quantum 
process is Markovian if and only if the dynamics is P-divisible. According to 
condition (\ref{P-divisibilty-condition}) P-divisibility implies that the rates 
(\ref{quantum-pauli-rates}) are indeed positive. Thus, we conclude that any 
quantum Markovian process which preserves the diagonal structure 
of quantum states in a fixed basis leads to a classical Markovian jump process 
describing transitions between the eigenstates of the density 
matrix.

\subsection{Alternative approaches}
\label{sec:altern-appr}

In Sec.~\ref{q-NM-info-flow} we have considered an approach to quantum
non-Markovianity based on the study of the dynamical behavior of the
distinguishability of states prepared by two distinct parties, Alice
and Bob. This viewpoint directly addresses the issue of the
experimental detection of quantum non Markovian processes, which can
be obtained by suitable quantum tomographic measurements. It further
highlights a strong connection between quantum non-Markovianity and
quantum information theory, or more specifically quantum information
processing.

Alternative approaches to the study of quantum non-Markovian dynamics
of open systems have also been recently proposed. We shall try to
briefly discuss those which are more closely related to the previous
study, without any claim to properly represent the vast literature on the
subject. For a review on recent results on quantum non-Markovianity see
\textcite{Rivas2014a}. The different approaches can to some extent be
grouped according to whether they propose different divisibility
properties of the quantum dynamical map
\cite{Wolf2008a,Rivas2010a,Hou2011a,Andersson2014a,Chruscinski2014a},
different quantifiers of the distinguishability between states
\cite{Vasile2011a,Chruscinski2012a,Lu2010a,Wissmann2013a,Dajka2011a} 
or the study of other
quantities, be they related to quantum information concepts or not,
which might exhibit both a monotonic and an oscillating behavior in
time \cite{Lu2010a,Luo2012a,Fanchini2014a,Haseli2014a,Bylicka2014a}.

\subsubsection{Quantum non-Markovianity and CP-divisibility}
\label{sec:quant-non-mark}

Let us first come back to the notion of divisible maps introduced in
Sec.~\ref{Div-time-local}, which we express in the form
\begin{equation}
  \label{eq:2}
  \Phi_{t,s}=\Phi_{t,\tau}\Phi_{\tau,s}, \qquad t\geq \tau\geq s \geq 0,
\end{equation}
where the maps have been defined according to
(\ref{two-parameter-family}). In the classical case considering the
matrices whose elements are given by the conditional transition
probabilities \eqref{def-T} according to
$(\Lambda_{t,s})_{xy}=T(x,t|y,s)$ the Chapman-Kolmogorov
equation takes the form
\begin{equation}\label{eq:3}
 \Lambda_{t,s}
 = \Lambda_{t,\tau}\Lambda_{\tau,s}, \qquad t\geq \tau\geq s \geq 0,
\end{equation}
where each $\Lambda_{t,s} $ is a stochastic matrix, which obviously
bears a strict relationship with \eqref{eq:2}, once one considers that
positive maps are natural quantum counterpart of classical
stochastic matrices.  Given the fact that in the classical case the
Markov condition \eqref{classical-Markov-condition} entails the
Chapman-Kolmogorov equation, and building on the conclusion drawn at
the end of Sec.~\ref{NM-quantum-regime}, i.e. that an
intrinsic characterization of memory effects has to be based on the
system's density operator only, whose dynamics is fully described in
terms of the time evolution maps, one is led to associate quantum
Markovianity with P-divisibility.  This observation reinforces the
result obtained in Sec.~\ref{gen-trace-distance-measure} relying on
the study of the information flow as quantified by the generalized trace
distance measure. Actually it can be shown that the P-divisibility
condition \eqref{eq:3} is equivalent to the monotonicity property of
the function
\begin{equation}
  \label{eq:4}
  K(P_1^{1}(t),P_1^{2}(t))=\sum_{x} |p_1 P_1^{1}(x,t)-p_2 P_1^{2}(x,t)|
\end{equation}
which provides a generalization of the Kolmogorov distance between two
classical probability distributions $P_1^{1}(t)$ and $P_1^{2}(t)$. Therefore these 
two equivalent properties capture the feature of a classical Markovian process at 
the level of the one-point probability distribution, which is all we are
looking for in order to obtain an intrinsic definition of quantum
non-Markovianity.

Relying on the fact that in the theory of open quantum systems
complete positivity suggests itself as a natural counterpart of positivity, as
an alternative approach to quantum non-Markovianity it has been
proposed to identify quantum Markovian dynamics with those dynamics
which are actually CP-divisible. This criterion was first suggested by
\textcite{Rivas2010a}. In order to check CP-divisibility of the time
evolution one actually needs to know the mathematical expression of
the transition maps $\Phi_{t,s}$ or of the infinitesimal generator
$\mathcal{K}_t$. Indeed given the maps $\Phi_{t,s}$ one can determine
their complete positivity studying the associated Choi matrices, while
as discussed in Sec.~\ref{Basic-Concepts} CP-divisibility corresponds
to positivity at all times of the rates $\gamma_i(t)$ appearing in \eqref{tcl-master}. 
In order to quantify non-Markovianity as described by this criterion, different
related measures have been introduced, relying on an estimate of the
violation of positivity of the Choi matrix \cite{Rivas2010a} or on various quantifiers
of the negativity of the rates \cite{Andersson2014a}.
In particular, \textcite{Rivas2010a} suggest to consider the following
measure of non-Markovianity based on CP-divisibility
\begin{equation}
\label{cpdiv}
\mathcal{N}_{\textrm{RHP}}(\Phi) = \int_0^{\infty} dt \, g(t)
\end{equation}
with
\begin{eqnarray}
 g(t) =
 \lim_{\epsilon\rightarrow0^+}\frac{|| \left( \Phi_{t+\epsilon,t} \otimes I \right)
 \left( | \Psi \rangle \langle \Psi | \right) ||-1}{\epsilon},
\end{eqnarray}
where $\ket{\Psi}$ is a maximally entangled state between open system and 
ancilla.  

A further feature which can be
considered in this context, and which has been proposed in
\textcite{Lorenzo2013a} as another
signature of non-Markovianity is the dynamical behavior of the volume
of the accessible states, according to their parametrization in the
Bloch representation, identifying non-Markovianity
with the growth of this volume, which also allows for a geometrical
description. This study can be performed e.g. upon knowledge of the
rates appearing in the time-local master equation, and appear to be a
strictly weaker requirement for Markovianity with respect to
P-divisibility \cite{Chruscinski2013a}.

To explain the relation between P-divisibility and CP-divisibility
suppose that Alice prepares states $\tilde{\rho}^1_S$ or
$\tilde{\rho}^2_S$ on an extended space
${\mathcal{H}}_S\otimes{\mathcal{H}}_S$ and sends the states through
the channel $\tilde{\Phi}_t = \Phi_t \otimes I$ to Bob.  The dynamical
map $\tilde{\Phi}_t$ is then Markovian, in the sense of P-divisibility,
if and only if $\Phi_t$ is CP-divisible. Note however that this
construction requires the use of a physically different system,
described by the tensor product space
${\mathcal{H}}_S\otimes{\mathcal{H}}_S$, which carries the information
transferred from Alice to Bob. More specifically, the maximization over
the initial states must allow for entangled states
$\tilde{\rho}^{1,2}_S$ which clearly explains the difference between
the definitions of non-Markovianity for $\Phi_t$ and $\tilde{\Phi}_t$.
This implies in particular that an experimental assessment of
non-Markovianity according to the criterion of CP-divisibility would
in general imply the capability to prepare entangled states and
perform tomographic measurements on the extended space.
Alternatively one should be able to perform process tomography instead of
state tomography, with an analogous scaling with respect to the
dimensionality of the system.

\subsubsection{Monotonicity of correlations and entropic quantities}
\label{sec:corr-monot}

Another line of thought about non-Markovianity of a quantum dynamics
is based on the study of the behavior in time of quantifiers of
correlations between the open system of interest and an ancilla
system. In this respect both entanglement \cite{Rivas2014a} and
quantum mutual information have been considered \cite{Luo2012a}.  In
this setting one considers an ancilla system and studies the behavior
in time of the entanglement or the mutual information of an initially
correlated joint system-ancilla state. Given the fact that both
quantities are non-increasing under the local action of a completely
positive trace preserving transformation, CP-divisibility of the time
evolution would lead to a monotonic decrease of these
quantities. Their non monotonic behavior can therefore be taken as a
signature or a definition of quantum non-Markovianity. Actually,
\textcite{Rivas2014a} have considered CP-divisibility as the distinguishing
feature of non-Markovianity, so that a revival of entanglement in the
course of time is just interpreted as a witness of
non-Markovianity. In \cite{Luo2012a} instead failure of monotonicity
in the loss of quantum mutual information is proposed as a new
definition of non-Markovianity. In both cases one considers as initial
state a maximally entangled state between system and ancilla. The
quantification of the effect is obtained by summing up the increases
over time of the considered quantity.  Still other approaches have
connected non-Markovianity with non monotonicity of the quantum
Fisher information, considered as a quantifier of the information
flowing between system and environment. In this case however the
information flow is not directly traced back to a distance on the
space of states \cite{Lu2010a}.

In the trace distance approach one studies the behavior of the
statistical distinguishability between states, and relying on the
introduction of the Helstrom matrix \cite{Chruscinski2011a} this
notion can be directly related to a divisibility property of the
quantum dynamical map. The generalization of the trace distance
approach also allows for a stronger connection to a quantum
information theory viewpoint. Indeed one can consider the quantum
dynamical maps as a collection of time dependent channels. In this
respect the natural question is how well Bob can recover information on
the state prepared by Alice performing a measurement after a given
time $t$. A proposal related to this viewpoint has been considered in
\cite{Bylicka2014a}, where two types of capacity of a quantum channel,
namely the so-called entanglement-assisted classical capacity and the
quantum capacity, have been considered. While the quantum data
processing inequality \cite{Nielsen2000} warrants monotonicity of
these quantities provided the time evolution is CP-divisible, lack of
this monotonicity has been taken as definition of non-Markovianity of
the dynamics.  The connection between non-Markovianity and the quantum
data processing inequality has also been recently studied by
\textcite{Buscemi-xxx}.

\section{Models and applications of non-Markovianity}

In this section we first address the description of simple
prototypical systems, for which one can exactly describe the
non-Markovian features of the dynamics according to the concepts and
measures introduced in Sec.~\ref{Definitions-Measures}. 
In particular we will stress how a
non-Markovian dynamics leads to a recovery of quantum features
according to the notion of information backflow.  In the second part
we present applications of the introduced non-Markovianity measures
to a number of more involved models illustrating how memory effects
allow to analyze their dynamics and can be related to characteristic properties 
of complex environments, thus suggesting to use the non-Markovianity of the
open system dynamics as a quantum probe for features of the environment.

\subsection{Prototypical model systems}\label{prot-model-sys}

We consider first a two-level system undergoing pure decoherence. 
This situation can be described by a time-convolutionless master equation of the form 
\eqref{tcl-master} with a single decay channel and allows to demonstrate in a simple 
manner how the properties of the environment and the system-environment coupling
strength influence the transition from a Markovian to a non-Markovian
dynamics for the open quantum system. A similar analysis is then
performed for a dynamics which also takes into account dissipative
effects, further pointing to the phenomenon of the failure of divisibility of
the quantum dynamical map. We further discuss a dynamics driven by
different decoherence channels, which allows to better discriminate
between different approaches to the description of non-Markovian behavior.
Finally, we present some results on the non-Markovian dynamics of the spin-boson model.

\subsubsection{Pure decoherence model}\label{sec:dephasing}

Let us start considering a microscopic model for a pure decoherence
dynamics which is amenable to an exact solution and describes a single
qubit interacting with a bosonic reservoir.
The total Hamiltonian
can be written as in Eq.~\eqref{Ham-total} with the system Hamiltonian $H_S$ and
the environmental Hamiltonian $H_E$ given by
\begin{equation}
  \label{eq:free}
  H_S=\frac 12 \omega_0 \sigma_z, \qquad 
  H_E=\sum_k \omega_k a_k^{\dagger} a_k,
\end{equation}
where $\omega_0$ is the energy difference between ground state
$\ket{0}$ and excited state $\ket{1}$ of the system, $\sigma_z$ denotes a
Pauli matrix, $a_k$ and $a_k^{\dagger}$ are annihilation and creation operators for the 
bosonic reservoir mode labelled by $k$ with frequency $\omega_k$, obeying the canonical
commutation relations $[a_k,a_{k'}^{\dagger}]=\delta_{kk'}$. The
interaction term is taken to be
\begin{equation}
  \label{eq:intz}
  H_I= \sum_k \sigma_z  \left( g_k a_k +  g_k^{\ast} a_k^{\dagger} \right)
\end{equation}
with coupling constants $g_k$. Considering a factorized initial condition
with the environment in a thermal state at inverse temperature $\beta$
the model can be exactly solved leading to a quantum dynamical map
$\Phi_t$ which leaves the populations invariant and modifies the
off-diagonal matrix elements according to
\begin{equation}\label{dephasing}
  \rho_{11}(t)=\rho_{11}(0), \qquad \rho_{10}(t)=G(t) \rho_{10}(0),
\end{equation}
where $\rho_{ij}(t)=\langle i | \rho_S(t) | j \rangle$ denote the
elements of the interaction picture density matrix
$\rho_{S}(t)$. The function $G(t)$ is often called decoherence
function and in the present case it is real since we are working in
the interaction picture. It can be expressed in the form
\begin{equation}
  \label{eq:6}
  G(t)=\exp\!\left[ -\!\!\int_0^{\infty} d\omega \, J(\omega)
    \coth\!\left(\! \frac{\beta\omega}{2} \!\right)\!
    \frac{1\!-\!\cos\left(\omega t\right)}{\omega^2}\right]\!,
\end{equation}
where we have introduced the spectral density $J(\omega)=\sum_k
|g_k|^2 \delta(\omega-\omega_k)$ which keeps track of the features of
the environment relevant for the reduced system description,
containing informations both on the environmental density of the modes
and on how strongly the system couples to each mode. It can be shown
that the interaction picture operator $\rho_S(t)$
obeys a master equation of the form \eqref{tcl-master} with a single
Lindblad operator,
\begin{equation}
\label{eq:dephase}
\frac{d}{dt}\rho_S(t) = \gamma (t) \left[  \sigma_z \rho_S(t)\sigma_z -\rho_S(t)\right],
\end{equation}
where the time dependent decay rate $\gamma (t)$ is related to the
decoherence function $G(t)$ by
\begin{equation}
\label{gammaG}
\gamma(t) = -\frac{1}{G(t)}\frac{d}{dt}  G(t).
\end{equation}
Note that for a generic microscopic or phenomenological decoherence model the 
decoherence function is in general a complex
quantity even in the interaction picture, and in this case the modulus
of the function should be considered in formula~\eqref{gammaG}.
The map $\Phi_t$ is completely positive since $G(t)\leq 1$ which is equivalent
to the positivity of the time integral of the decay rate $\Gamma(t)=\int_0^t dt' \, \gamma(t')$.
In order to discuss the non-Markovianity of the obtained time evolution as discussed in
Sec.~\ref{q-NM-info-flow} and Sec.~\ref{sec:quant-non-mark} we
consider the divisibility property of the time evolution. The master
equation \eqref{eq:dephase} has a single channel with decay rate
$\gamma (t)$ and thanks to the strict positivity of the decoherence function
$\Phi_t^{-1}$ always exists. However, according to the criterion given by
Eq.~\eqref{P-divisibilty-condition}, P-divisibility fails at times when the
decoherence rate becomes negative, which is exactly when
CP-divisibility is lost. This is generally true when one has a
master equation in the time-local form \eqref{tcl-master} with a
single Lindblad operator, since in this case the condition for
CP-divisibility, given by the requirement of positivity of the decay
rate $\gamma(t)$, coincides with the condition
\eqref{P-divisibilty-condition} for P-divisibility.

The trace distance between time evolved states corresponding to the
initial conditions $\rho_S^{1,2}$ is given according to
\eqref{trace-distance} by
\begin{equation}
  \label{eq:7}
  D\left(\Phi_t\rho^1_S,\Phi_t\rho^2_S\right)=\sqrt{a^2+G^2(t)|b|^2},
\end{equation}
where $a=\rho_{11}^1-\rho_{11}^2$ and $b=\rho_{10}^1-\rho_{10}^2$
denote the difference of the populations and of the coherences of the
initial states respectively. Its time derivative corresponding to the
quantity \eqref{trace-distance-deriv} in terms of which the
non-Markovianity measure can be constructed then reads
\begin{equation}
  \label{eq:8}
  \sigma(t)=\frac{G(t)|b|^2}{\sqrt{a^2 + G^2(t)|b|^2}}
 \frac{d}{dt}G(t).
\end{equation}
The non-monotonicity in the behavior of the trace distance which is
used as a criterion for non-Markovianity in Sec.~\ref{def-NM-measure}
is therefore determined by the non monotonic behavior of the
decoherence function. In order to evaluate the measure
Eq.~\eqref{NM-measure} one has to maximize over all pairs of
initial states. As shown by \textcite{Wissmann2012a} and
discussed in Sec.~\ref{def-NM-measure} the latter
can be restricted to orthogonal state pairs. For the case at hand it follows from
Eq.~\eqref{eq:8} that the maximum is obtained for antipodal points on
the equator of the Bloch sphere, so that $a=0$ and $|b|=1$. The
non-Markovianity measure \eqref{NM-measure} is then given by
\begin{equation}
\label{nm}
\mathcal{N}(\Phi) = \sum_k \left[ G (t_k^f)-G \left ( t_k^i\right) \right],
\end{equation}
where $t_k^i$ and $t_k^f$ denote initial and final time point of the
$k$-th interval in which $G(t)$ increases. Since the increase of
$G(t)$ coincides with the negativity of the decoherence rate according to
Eq.~\eqref{gammaG}, the growth of the trace distance is in this case
equivalent to breaking both P-divisibility and CP-divisibility. Moreover,
the generalized trace distance measure considered in
Sec.~\ref{gen-trace-distance-measure} leads to the same expression
\eqref{nm} for non-Markovianity. 

The measure for non-Markovianity of Eq.~\eqref{cpdiv} based on
CP-divisibility takes in this case the simple expression
\begin{equation}
\label{eq:13}
\mathcal{N}_{\textrm{RHP}}(\Phi) =-2 \int_{\gamma<0} dt~\gamma(t),
\end{equation}
which is the integral of the decoherence rate in the time intervals in
which it becomes negative.
An example of an experimental realization of a pure decoherence model is
considered in Sec.~\ref{sec:contr-quant-memory}, where the
decoherence function is determined by the interaction between
the polarization and the frequency degrees of freedom of a photon. In this case
the decoherence function is, in general, complex valued, so that the previous formulae
hold with $G(t)$ replaced by $|G(t)|$. 
The geometric measure discussed in Sec.~\ref{sec:quant-non-mark}, which connects
non-Markovian behavior with the growth in time of the volume of
accessible states, also leads to the same signature for
non-Markovianity: The volume can again be identified with
the decoherence function $G(t)$ and therefore as soon as $G(t)$
increases the volume grows. 

As we have discussed, in a simple decoherence model non-Markovianity can be
identified with the revival of coherences of the open system,
corresponding to a backflow of information from the environment to the
system. In this situation the time-local master equation
\eqref{eq:dephase} describing the dynamics exhibits just a single
channel, so that all discussed criteria of non-Markovianity do coincide.

\subsubsection{Two-level system in a dissipative environment}\label{sec:dissipation}

We now consider an example of dissipative dynamics in which the
system-environment interaction influences both coherences and
populations of a two-state system.
The environment is still taken to be a bosonic bath, so that the free
contributions to the Hamiltonian are again given by
Eq.~\eqref{eq:free}, but one considers an interaction term in rotating
wave approximation given by
\begin{equation}
\label{eq:rwa}
H_I= \sum_k \left( g_k \sigma_+ a_k +  g_k^{\ast} \sigma_- a_k^{\dagger} \right),
\end{equation}
where $\sigma_-= |0\rangle \langle 1 |$ and $\sigma_+= |1\rangle
\langle 0 |$ are the lowering and raising operators of the
system. Such an interaction describes, e.g., a two-level atom in a
lossy cavity. For a factorized initial state with the environment in
the vacuum state, this model is again exactly solvable, thanks to the conservation 
of the number of excitations. The quantum dynamical map
$\Phi_t$ transforms populations and coherences according to
\begin{equation} \label{eq:9}
  \rho_{11}(t)=|G(t)|^2 \rho_{11}(0), \qquad \rho_{10}(t)=G(t) \rho_{10}(0),
\end{equation}
where the decoherence function is a complex function determined by the
spectral density $J(\omega)$ of the model. In particular, denoting by
$f(t-t_1)$ the two-point correlation function of the environment
corresponding to the Fourier transform of the spectral density, the
function $G(t)$ is determined as the solution of the integral equation 
 \begin{equation}
\frac{d}{dt} G(t) = -\int dt_1\, f(t-t_1) G(t_1)
\end{equation}
with initial condition $G(0)=1$.
Also in this case it is possible to write down the exact master
equation obeyed by the density matrix 
$\rho_S(t)$ of the system in the form of Eq.~\eqref{tcl-master},
which in the interaction picture reads \cite{Breuer1999b}
\begin{eqnarray}
\label{eq:merwa}
\frac{d}{dt}\rho_S(t) =&-&\frac{\displaystyle i}{\displaystyle 4} S(t) \left[ \sigma_z,
  \rho_S\right] \\  \nonumber
 &+& \gamma (t) \left[  \sigma_- \rho_S(t)\sigma_+ - \frac{1}{2}\left\{\sigma_+\sigma_-,\rho_S(t)\right\}\right],
\end{eqnarray}
where the time dependent Lamb shift is given by $S(t)=-2 \Im
({\dot{G}(t)}/{G(t)})$, and the decay rate can be written as $\gamma(t) = -2{\Re}
({\dot{G}(t)}/{G(t)})$, or equivalently
\begin{equation}\label{eq:12}
\gamma(t) =-\frac{2}{|G(t)|} \frac{d}{dt}|G(t)|.
\end{equation}
The analysis of non-Markovianity in the present model closely follows
the discussion in \ref{sec:dephasing} due to the crucial fact that
the master equation \eqref{eq:merwa} still has a single Lindblad
operator. The expression of the trace distance is determined by the
decoherence function and, with the same notation as in \eqref{eq:7}, reads
\begin{equation}
  \label{eq:5}
  D\left(\Phi_t\rho^1_S,\Phi_t\rho^2_S\right)=|G(t)|\sqrt{|G(t)|^2 a^2+|b|^2},
\end{equation}
so that one has the time derivative
\begin{equation}
  \label{eq:10}
  \sigma(t)=\frac{2|G(t)|^2a^2 + |b|^2}{\sqrt{|G(t)|^2a^2 + |b|^2}}
 \frac{d}{dt}|G(t)|.
\end{equation}
Complete positivity of the map is ensured by positivity of the
integrated decay rate $\Gamma(t)$, while the trace distance shows a
non monotonic behavior when the modulus of the decoherence function
grows with time. 

A typical expression for the spectral density is given by a Lorentzian 
\begin{equation}
  \label{eq:11}
  J(\omega) = \gamma_0 \lambda^2 / 2 \pi \left[ (\omega_0+\Delta- \omega)^2 + \lambda^2 
  \right]
\end{equation}
of width $\lambda$ and centered at a frequency detuned from the atomic
frequency $\omega_0$ by an amount $\Delta$, while the rate $\gamma_0$
quantifies the strength of the system-environment coupling. Let us first discuss the case
when the qubit is in resonance with the central frequency of the spectral
density, so that one has a vanishing detuning, $\Delta =0$. The decoherence function 
then takes the form
\begin{equation}
 G(t) = e^{-\lambda t/2}\left[\cosh\left(\frac{dt}{2}\right)
 +\frac{\lambda}{d}\sinh\left(\frac{dt}{2} \right)\right],
\end{equation}
where $d=\sqrt{\lambda^2-2\gamma_0\lambda}$. For weak couplings,
corresponding to $\gamma_0<\lambda/2$, the decoherence function $G(t)$
is a real, monotonically decreasing function so that \eqref{eq:10} is
always negative and the dynamics is Markovian. In this case, as
discussed in Sec.~\ref{Div-time-local}, the map is invertible and since
the time-convolutionless form of the master
equation has just a single channel, it is at the same time P-divisible
and CP-divisible. Note that in the limit $\gamma_0 \ll \lambda/2$ the
rate $\gamma(t)$ becomes time-independent, $\gamma=\gamma_0$, such that
the master equation \eqref{eq:merwa} is of Lindblad form and leads to a Markovian 
semigroup. However, for larger values of the coupling between
system and environment, namely for $\gamma_0>\lambda/2$, the function
$|G(t)|$ displays an oscillatory behavior leading to a non-monotonic time evolution
of the trace distance and, hence, to non-Markovian dynamics. Thus, the
trace distance measure of Eq.~\eqref{NM-measure} is zero for $\gamma_0<\lambda/2$
and starts at the threshold $\gamma_0=\lambda/2$ to increase continuously to positive values  
\cite{Laine2010a}. It is interesting to note that
the transition point $\gamma_0=\lambda/2$ exactly coincides with the
point where the perturbation expansion of the time-convolutionless master equation
\eqref{eq:merwa} breaks down. We also remark that for this model optimal state pairs still
correspond to antipodal points of the equator of the Bloch sphere \cite{Xu2010b}, and  that
the criterion for non-Markovianity based on the Helstrom matrix
considered in Sec.~\ref{gen-trace-distance-measure} leads to the same results.

In the parameter regime $\gamma_0>\lambda/2$ the map $\Phi_t$ defined by
\eqref{eq:9} is no longer invertible for all times due to the existence of zeros of the 
decoherence function. Thus, the model provides an example for a family of maps 
lying in the lower (red) region depicted in Fig.~\ref{fig:scheme}. Physically, the two-level 
system reaches the ground state before memory effects revive the coherences
and the population of the upper state. Indeed, also in this case
non-Markovianity can be traced back to the revival of populations and coherences, 
corresponding to a backflow of information from the environment to the system.
Since the inverse of $\Phi_t$ does not exist for all times, the criterion based
on CP-divisibility cannot strictly speaking be applied. This is
reflected by the fact that the corresponding measure, taking in this case again the 
expression \eqref{eq:13}, jumps from zero to infinity.  

In the non-resonant case, i.e., for nonzero detuning, one can
observe a transition from Markovian to non-Markovian dynamics for a
fixed value of the coupling in the weak coupling regime by increasing 
the detuning $\Delta$. In this case
the map is invertible for all times, since the decay rate $\gamma(t)$ 
is finite for all times, so that also the measure \eqref{cpdiv} remains well
defined. Again, the regions of non-Markovianity coincide for all
approaches since the constraint in order to have CP-divisibility or
P-divisibility is the same. The pair of states maximizing the expression
of the measure \eqref{NM-measure} are however now given by the
projections onto the excited and the ground state. This shows how the optimal
pair, which is always given by orthogonal states, does depend on the actual
expression for the decoherence function. For this case the derivative
of the trace distance \eqref{eq:10} takes the simple
form $\sigma(t)= -\gamma(t) \exp [-\Gamma(t)]$ with
$\Gamma (t) = \int_0^t \gamma(t')~dt' $, which puts into evidence the
direct connection between the direction of information flow and the
sign of the decay rate.

\subsubsection{Single qubit with multiple decoherence channels}

In the preceding examples we have shown that for a master equation in the 
time-convolutionless form \eqref{tcl-master} with a single Lindblad operator $A(t)$, 
all the different considered criteria for non-Markovianity coincide.  It is therefore of interest to 
discuss an example in which one has multiple decoherence channels with generally different 
rates. As we shall shortly see, such model allows to discriminate among the different definitions 
of non-Markovian dynamics. To this end, we take a phenomenological approach and following 
\textcite{Chruscinski2013a,Chruscinski2014a,Vacchini2012a} we consider for a two-level system 
the master equation
\begin{equation}
 \label{eq:ru}
 \frac{d}{dt}\rho_S(t) 
 = \frac 12 \sum_{i=1}^{3} \gamma_i(t) \left[  \sigma_i \rho_S(t)\sigma_i -\rho_S(t)\right],
\end{equation}
where $\sigma_i$ with $i=x,y,z$ denote the three Pauli
operators. Considering, e.g., the decoherence dynamics of a spin-$\frac{1}{2}$ particle in
a complex environment the rates $\gamma_i(t)$ would correspond to
generally time dependent transversal and longitudinal decoherence
rates.  The dynamical map corresponding to \eqref{eq:ru} can be
exactly worked out and is given by the random unitary dynamics 
\begin{equation} \label{eq:16}
 \Phi_t(\rho_S) = \sum_{i=0}^3 p_i(t) \sigma_i \rho_S \sigma_i \, ,
\end{equation}
where $\sigma_0$ denotes the identity operator and the coefficients
$p_i(t)$ summing up to one are determined from the decoherence rates.
Introducing the quantities $A_{ij}(t)= \exp[-(\Gamma_i(t) +
\Gamma_j(t))]$, where as usual we have denoted by $\Gamma_k(t)$ the
time integral of the $k$-th decay rate, the coefficients $p_i(t)$ take
the explicit form $p_{0,1}= (1/4) [1\pm A_{12}(t) \pm A_{13}(t) +
A_{23}(t)]$ and $p_{2,3}= (1/4) [1\mp A_{12}(t) \pm A_{13}(t) -
A_{23}(t)]$. These equations show that the dynamical map actually
depends on the sum of the integrals of the decay rates. According to
its explicit expression the map $\Phi_t$ is completely positive
provided the coefficients $p_i(t)$ remain positive, 
in which case they can
be interpreted as a probability distribution and thus characterize the
random unitary dynamics \eqref{eq:16}. Note that this can be
the case even if a decoherence rate stays negative at all times as in 
\textcite{Andersson2014a,Vacchini2011a}. 

As explained in Sec.~\ref{Div-time-local}, the map is CP-divisible when all of the
decoherence rates remain positive for all $t\geqslant 0$, so that as
soon as at least one of the rates becomes negative the measure
\eqref{cpdiv} becomes positive indicating a non-Markovian
behavior. However, the condition \eqref{P-divisibilty-condition} for having P-divisibility 
is weaker. Indeed, in order to have P-divisible dynamics only the sum of all pairs of distinct 
decoherence rates has to remain positive, i.e., $\gamma_i(t)+\gamma_j(t)\geqslant0$ 
for all $j\not =i$.
The same constraint warrants monotonicity in time of the behavior of the
trace distance, so that also in this case the measure
\eqref{NM-measure} of non-Markovianity and its generalized version
\eqref{NM-extended-measure} based on the Helstrom matrix and corresponding to 
P-divisibility lead to the same result. 
In the characterization of non-Markovianity based on the backflow of 
information from the environment to the system memory effects 
therefore only appear whenever the sum of at least a 
pair of decoherence rates becomes negative. 
To better understand, how this fact can be related to the dynamics of
the system, one can notice \cite{Chruscinski2014a} that the Bloch vector components 
$\langle\sigma_i(t)\rangle$, according to Eq.~\eqref{eq:ru}, obey the equation
\begin{equation}
  \label{eq:15}
  \frac{d}{dt}\langle
\sigma_i(t)\rangle = -\frac{1}{T_i(t)} \langle
\sigma_i(t)\rangle,
\end{equation}
where the relaxation times are given by $T_i(t) = [\gamma_j(t) +
\gamma_k(t)]^{-1}$ (with all three indices taken to be
distinct) and correspond to experimentally measurable quantities.
Appearance of non-Markovianity and therefore failure of
P-divisibility is thus connected to negative relaxation rates,
corresponding to a rebuild of quantum coherences.
For the present model one can also easily express the condition
leading to a growth of the volume of accessible states within the Bloch sphere. 
In fact, this volume of accessible states is proportional to
$\exp [-\sum_{i=1}^3\Gamma_i(t)]$. As a consequence, the dynamics according to the
geometric criterion by \textcite{Lorenzo2013a} is Markovian if and only
if $\sum_{i=1}^3\gamma_i (t)\geqslant 0$ for $t\geqslant0$, which is a
strictly weaker criterion with respect to either P or CP-divisibility.

While in the examples considered above the non-Markovianity measure
\eqref{NM-measure} based on the trace distance and its generalization
\eqref{NM-extended-measure} based on the Helstrom matrix agree in the
indication of non-Markovian dynamics, a situation in which these
criteria are actually different has been considered by \textcite{Chruscinski2011a}.

\subsubsection{Spin-boson model}\label{sec:spin-boson}

\begin{figure}[t]
\includegraphics[width=0.40\textwidth]{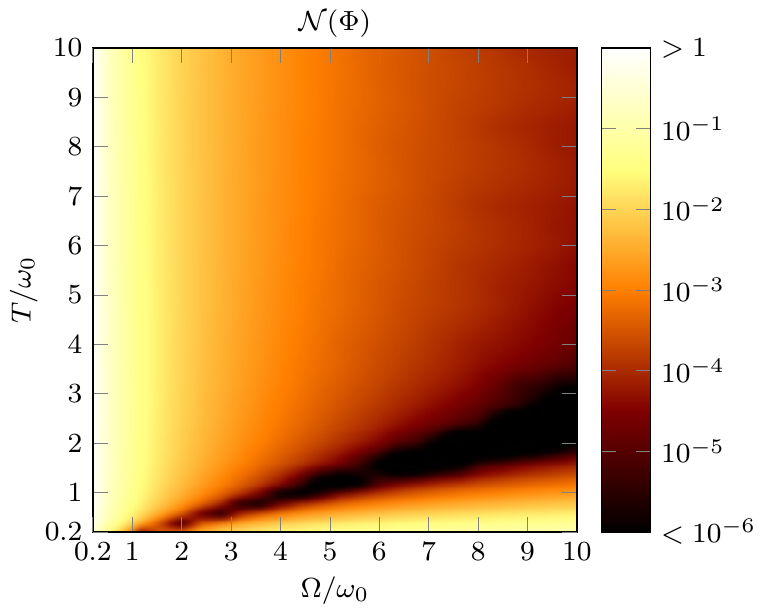}
\caption{(Color online) The non-Markovianity measure $\mathcal{N}(\Phi)$ 
defined by Eq.~\eqref{NM-measure} for the spin-boson model as a
function of temperature $T$ and cutoff frequency $\Omega$ in units of the
transition frequency $\omega_0$ \cite{Clos2012a}.}
\label{fig:spin_boson}
\end{figure}

Finally, we briefly discuss a further paradigmatic model for dissipative quantum
dynamics with many applications, namely the spin-boson model \cite{Leggett1987}.
The system and the environmental Hamiltonian are again given by 
Eq.~\eqref{eq:free}, while the interaction Hamiltonian has the
non-rotating-wave structure
\begin{equation}\label{H_I_spin_boson}
 H_I = \sum_k \sigma_x \left( g_k a_k +  g_k^{\ast} a_k^{\dagger} \right).
\end{equation}
Figure \ref{fig:spin_boson} shows the non-Markovianity measure 
$\mathcal{N}(\Phi)$ for this model obtained from numerical 
simulations of the corresponding second-order time-convolutionless master 
equation, using an Ohmic spectral density of Lorentz-Drude shape with cutoff 
frequency $\Omega$, reservoir temperature $T$ and a fixed small coupling 
strength of size $\gamma=0.1\omega_0$ \cite{Clos2012a}. 
As can be seen from the figure the dynamics is strongly non-Markovian both for
small cutoff frequencies and for small temperatures (note the logarithmic scale 
of the color bar). The emergence of a Markovian region within the non-Markovian
regime for small cutoffs and temperatures can be understood in terms of a
resonance between the transition frequency and the maximum of an effective,
temperature-dependent spectral density of the environmental modes.

\subsection{Applications of quantum non-Markovianity}

In Sec.~\ref{prot-model-sys} we have illustrated quantum non-Markovian behavior by 
means of simple model systems. The goal of the present section is to demonstrate that 
memory effects and their control also open new perspectives for applications. In fact, 
understanding various aspects of non-Markovianity makes it possible to 
develop more refined tools for reservoir engineering and to use memory effects as an 
indicator for the presence of typical quantum features. Moreover, it becomes possible 
to develop schemes where a small open system is used as a quantum probe
detecting characteristic properties of the complex environment it is interacting with. 
Here, we present a few examples from the recent literature illustrating these points.

\subsubsection{Analysis and control of non-Markovian dynamics}

We first consider the model of a spin-$\frac{1}{2}$ particle with spin operator 
$\mathbf{s}_0$ coupled to a chain of $N$ spin-$\frac{1}{2}$ particles with spin operators 
$\mathbf{s}_n$ ($n=1,2,\ldots,N$), which has been investigated by \textcite{Apollaro2011a}.
The system Hamiltonian is given by $H_S=-2h_{0}s^z_{0}$, where $h_0$ denotes a
local field acting on the system spin $\mathbf{s}_0$. The environmental Hamiltonian 
represents an $XX$-Heisenberg spin chain with nearest neighbor interactions of strength 
$J$ in a transverse magnetic field $h$:
\begin{equation}
 H_E = -2J\sum_{n=1}^{N-1} \left( s^x_{n} s^x_{n+1} + s^y_{n} s^y_{n+1} \right)
 -2h\sum_{n=1}^{N} s_{n}^z.
\end{equation}
The interaction Hamiltonian takes the form
\begin{equation}
 H_I = -2J_0 \left( s^x_{0}s^x_{1}+s^y_{0}s^y_{1} \right),
\end{equation}
describing an energy-exchange interaction between the system spin $\mathbf{s}_0$ 
and the first spin $\mathbf{s}_1$ of the chain with coupling strength $J_0$. This
model leads to a rich dynamical behavior of the system spin and allows to demonstrate 
not only the impact of the system-environment interaction, but also how the open system 
dynamics changes when manipulating the interactions between the constituents of the 
environment and their local properties. 

It can be shown analytically that optimal state pairs for the trace distance measure 
correspond to antipodal points on the equator of the surface of the Bloch sphere.
Moreover, when the local fields of the system and environmental sites are equal, 
$h=h_0$, one can derive an analytic expression for the rate of change $\sigma(t)$ of the 
trace distance [see Eq.~\eqref{trace-distance-deriv}] for optimal state pairs:
\begin{equation}
 \sigma(t) = -(2/{t}) \, \text{sgn} [{\cal J}_1(2t)] \, {\cal J}_2(2t),
\end{equation}
where sgn is the sign function, and ${\cal J}_1$ and ${\cal J}_2$ denote the Bessel 
functions of order 1 and 2, respectively. In the physically more interesting case
$h \neq h_0$ the trace distance measure $\mathcal{N}(\Phi)$ 
[see Eq.~\eqref{NM-measure}] has to be determined numerically. 
Let us first consider the case, where the system local field is zero, $h_0=0$, 
the couplings between the spins are equal, $J=J_0$, and we tune the ratio $h/J$ 
of the local field with respect to the spin-spin interaction strength. It turns out that the open 
system is then Markovian only at the point $h/J=1/2$, i.e., when the strength of the local 
field is half of the coupling between the spins. For all other values of $h/J$ the open system 
dynamics displays memory effects with $\mathcal{N}(\Phi)>0$. Thus, the point of 
Markovianity separates two regions where memory effects are present.
This behavior persists for $h_0\neq 0$: Denoting the detuning between the local fields by
$\delta = h - h_0$ and still keeping $J=J_0$, the point of Markovianity occurs at 
$\delta h/J = 1/2$. On both sides of this point the system again exhibits memory effects,
but interestingly their properties are different. For $\delta h/J <1/2$ the system approaches 
a steady state which is independent of the initial state. On the other hand, for 
$\delta h/J >1/2$ the trace distance does not decay to zero asymptotically in time.
This feature can be interpreted as information trapping since it implies that the 
distinguishability of states does not vanish asymptotically. We also remark that the
origin of memory effects and the differences in the dynamical behavior can be 
understood in more detail by studying the spectrum of the total Hamiltonian
\cite{Apollaro2011a}.  
 
As our second example we discuss applications to a physical system which has been 
studied extensively during the last twenty years, namely Bose-Einstein condensates (BECs) 
which can be manipulated with high precision by current technologies and thus provide 
means for environment engineering \cite{Pitaevskii2003,Pethick2008,Dalfovo1999a}. 
Following \textcite{Haikka2011a} we consider an impurity-BEC 
system, where the Markovian to non-Markovian transition can be controlled by tuning the 
properties of the BEC. An impurity atom is trapped within a double-well potential,
where the left ($|L\rangle$) and right ($|R\rangle$) states represent the two qubit states. 
The BEC, which constitutes the environment of the qubit, is trapped in a harmonic potential. 
The Hamiltonian for the total impurity-BEC system reads
\begin{equation}
 H = \sum_\mathbf{ k} E_\mathbf{ k} c_\mathbf{ k}^\dagger c_\mathbf{ k}
 + \sum_\mathbf{ k}(\xi_\mathbf{ k}c_\mathbf{ k}^ \dagger 
 + \xi_\mathbf{ k}^*c_\mathbf{ k})
 +\sum_\mathbf{ k} \sigma_z (g_\mathbf{ k}c_\mathbf{ k}^ \dagger 
 +g_\mathbf{ k}^*c_\mathbf{ k}).
\end{equation}
Here, $E_\mathbf{ k}$ denotes the energy of the Bogoliubov mode $c_\mathbf{ k}$ of the 
condensate, $\sigma_z=|R\rangle\langle R|-|L\rangle\langle L|$, while $g_\mathbf{k}$ 
and $\xi_\mathbf{k}$ describe the impurity-BEC and the intra-BEC couplings, respectively.

For a background BEC at zero temperature the open qubit dynamics is described by a 
dephasing master equation of the form of Eq.~\eqref{eq:dephase} and it turns out that
the decoherence rate $\gamma(t)$ can be tuned by changing the inter-well distance, 
and  by controlling the dimensionality and the interaction strength (scattering length) of the 
BEC. The detailed studies demonstrate that for a 3D gas, an increase of the
well-separation and of the intra-environment interaction leads to an increase of the
non-Markovianity measure for the impurity dynamics. Moreover, the dimensionality of the 
environment influences the emergence of memory effects. Studying the 
Markovian to non-Markovian  transition as a function of the intra-environment interaction 
strength shows that a 3D environment is the most sensitive. 
The Markovian -- non-Markovian crossover occurs in 3D for weaker values of the 
scattering length than in 1D and 2D environments, and a 1D environment requires the 
largest scattering length for the crossover. This behavior can be ultimately traced back to the 
question of how the dimensionality of the environment influences the spectral 
density which governs the open system dynamics. In general, for small frequencies the 
spectral density shows a power law behavior, $J(\omega) \propto \omega^s$.
The spectral density is called sub-Ohmic, Ohmic, and super-Ohmic for 
$s<1$, $s=1$, and $s>1$, respectively. For a 3D gas, even without interactions 
within the gas, the spectral density has super-Ohmic character.
Introducing interactions within the environment then makes this character stronger and 
thereby the 3D gas is most sensitive to the Markovian to non-Markovian transition. In 
contrast, for the 1D gas with increasing scattering length the spectral density first changes 
from sub-Ohmic to Ohmic and finally to super-Ohmic which then allows the appearance of 
memory effects.

\subsubsection{Open systems as non-Markovian quantum probes}

In addition to detecting, quantifying and controlling memory effects, it turns out to be
fruitful to ask whether the presence or absence of such effects in the dynamics of 
an open quantum system allows to obtain important information about characteristic
features of a complex environment the open system is interacting with. 
To illustrate this point we present a physical scenario, where Markovian behavior indicates 
the presence of a quantum phase transition in a spin environment \cite{Sachdev2011}.
Consider a system qubit, with states $|g\rangle$ and $|e\rangle$, which interacts with an
environment described by a one-dimensional Ising model in a transverse field 
\cite{Quan2006}.
The Hamiltonians of the environment and the qubit-environment interaction are given by
\begin{eqnarray}
H_E&=&-J\sum_{j}\left({\sigma^{z}_{j}}{\sigma^{z}_{j+1}}+\lambda{\sigma^{x}_{j}}\right), \\
H_I&=&-J\delta|e\rangle\langle e|\sum_{j}\sigma^{x}_j, \label{H-int-Ising}
\end{eqnarray}
where $J$ is a microscopic energy scale, $\lambda$ a dimensionless coupling
constant describing the strength of the transverse field, and $\sigma_j^{x,z}$ are Pauli spin 
operators. The system qubit is coupled with strength $\delta$ to all environmental spins.
The model yields pure dephasing dynamics described by a 
decoherence function $G(t)$ which, for a pure environmental initial state $|\Phi\rangle$, is 
given by
\begin{equation}\label{G-Ising}
 G(t)=\langle\Phi|e^{i H_{g}t}e^{-i H_{e}t}|\Phi\rangle,
\end{equation} 
where the effective environmental Hamiltonians are 
$H_{\alpha} = H_{E} +\bra{\alpha}H_{I}\ket{\alpha}$ with $\alpha=g,e$. We note
that according to Eq.~\eqref{G-Ising} the decoherence function is directly linked to the
concept of the Loschmidt echo $L(t)$ \cite{Cucchietti2003} which characterizes how the 
environment responds to perturbations by the system
\cite{Peres1984,Gorin2006,Jalabert2008,Jalabert2004}. In fact, we have $L(t)=|G(t)|^{2}$ 
and, hence,
there is also a direct connection between the Loschmidt echo and the evolution of
the trace distance for optimal state pairs, i.e. $D(\rho_S^1(t), \rho_S^2(t))= \sqrt{L(t)}$. 

\begin{figure}[t]
\includegraphics[width=0.35\textwidth]{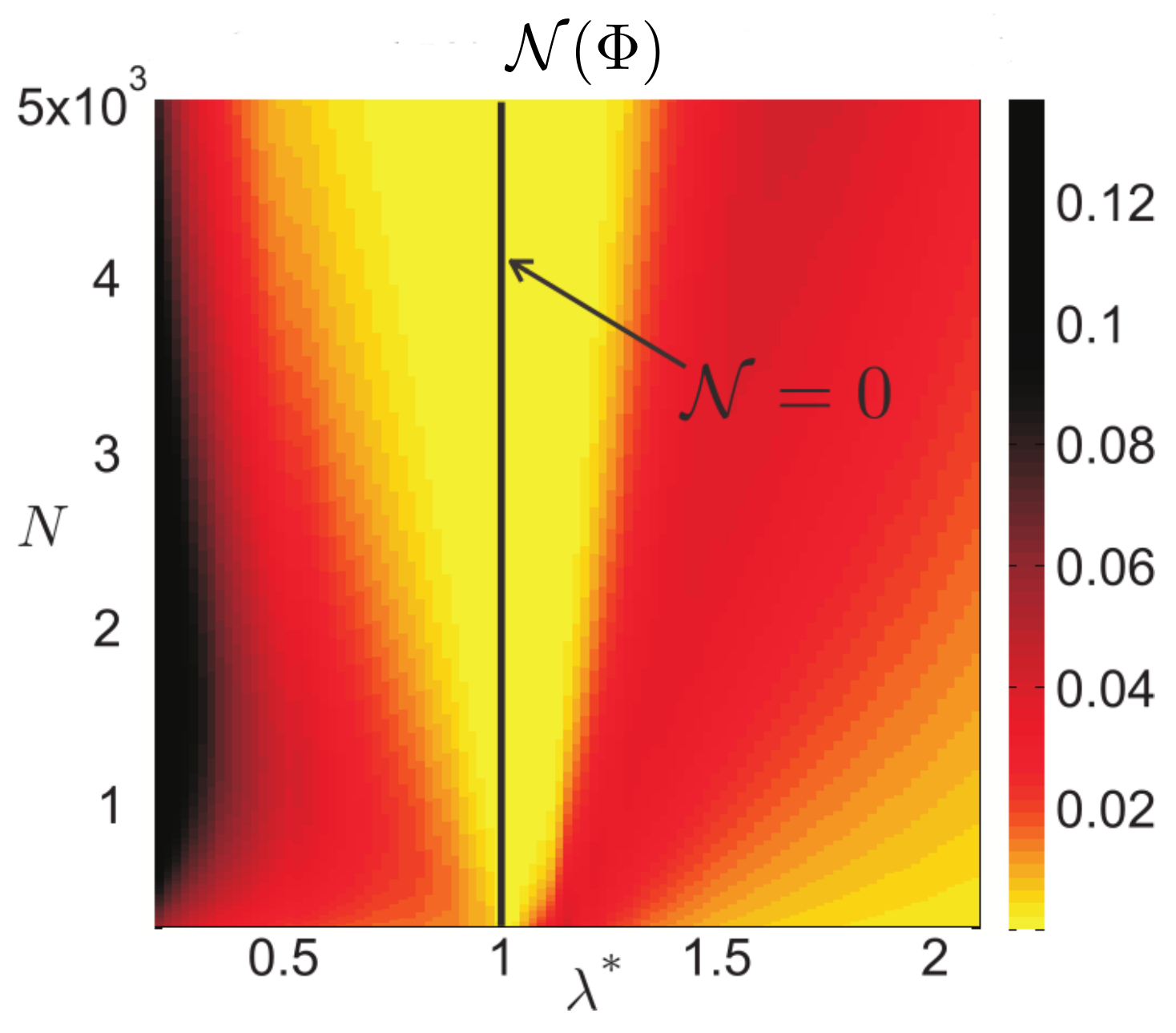}
\caption{(Color online) The non-Markovianity measure $\mathcal{N}(\Phi)$ of 
Eq.~\eqref{NM-measure} for a qubit coupled to a one-dimensional Ising chain in a 
transverse field as a function of field strength $\lambda^{\ast}$ and
particle number $N$. The measure vanishes along the black line and is nonzero
everywhere else \cite{Haikka2012a}.}
\label{trans}
\end{figure}

The system-environment interaction \eqref{H-int-Ising} leads to the effective
transverse field $\lambda^{\ast}=\lambda+\delta$ when the system is in the state 
$|e\rangle$. At the critical point of the spin environment ($\lambda^{\ast}=1$) the 
spin-spin interaction and the external field are of the same size
and the environment is most sensitive to external perturbations. This is reflected in a quite 
remarkable manner in the dynamics of the system qubit as can be seen from 
Fig.~\ref{trans} \cite{Haikka2012a}.  
At the point $\lambda^{\ast}=1$ the system qubit experiences strong 
decoherence. As a matter of fact, it turns out that the dynamics of the system qubit always 
shows non-Markovian behavior except at the critical point of the environment where the 
dynamics is Markovian. Therefore, the system qubit acts as a probe for the spin 
environment and Markovian dynamics represents a reliable indicator of environment 
criticality. It is important to note that even though, strictly speaking, quantum phase 
transitions require to take the thermodynamic limit, the above results hold for any number of 
environmental spins.

\section{Impact of correlations and experimental realizations}
In this section we study the effect of two fundamental kinds of correlations 
on the non-Markovian behavior of open systems, namely correlations
between the open system and its environment and correlations within
a composite environment. We conclude the section by giving an 
overview of the experimental status of non-Markovian open systems.

\subsection{Initial system-environment correlations}

\subsubsection{Initial correlations and dynamical maps}
Up to this point we have discussed the theory of open quantum systems
via the concept of a dynamical map, which in Sec.~\ref{open-sys-dyn-maps} 
we wrote down in terms of the initial state of
the environment and of the total Hamiltonian. In doing so we assumed that
the open system and the environment are independent at the initial
time of preparation [see Eqs.~\eqref{Product-initial-state} and
\eqref{Rho-S-repr}]. However, experimentalists often study fast
processes in strongly coupled systems, where this independence is
rarely the case, as originally pointed out by
\textcite{Pechukas1994a}. Going beyond this approximation brings us up
against the problem of describing the dynamical properties of open quantum 
systems without dynamical maps.

The theory of completely positive maps plays a crucial role in many fields of quantum physics 
and thus a general map description for open systems, even in the presence of initial 
correlations, would be very useful. Thus, naturally, a lot of effort has been put in 
trying to extend the formalism along these lines \cite{Alicki1995a, 
Stelmachovic2001a, Shabani2009a}. In this framework the aim has been to 
determine under which conditions the system dynamics can be described by
means of completely positive maps if initial correlations are present. Unfortunately, no generally 
acknowledged answer to this question has been found and the topic is still under 
discussion \cite{Shabani2009a, Rodriguez2010a, Brodutch2013a}.

As useful as a description in terms of completely positive maps would be, one may as well ask the 
question: How do the initial correlations influence the dynamics? Especially, since 
an experimentalist is often restricted to looking only at a subset of system states, 
he would want to conclude something about the system based on the dynamics of 
this subset. It is not possible to construct a map, but perhaps something else 
could be concluded. In the following we present an approach, in which the 
dynamics of information flow between the system and the environment is studied. 
We will see that the initial correlations modify the information flow and 
consequently can be witnessed from the dynamical features of the open system.

Let us have a closer look at the dynamics of the information inside the open 
system $\mathcal{I}_{\rm int}(t)$, defined in Eq.~\eqref{eq:1}, for a general open 
system, where initial correlations could be present. Suppose there are two 
possible preparations giving rise to distinct system-environment initial states 
$\rho_{SE}^1(0)$ and $\rho_{SE}^2(0)$. Since the total system and the 
environment evolve under unitary dynamics, we have
\begin{equation}\label{initial-inequality-1}
 \mathcal{I}_{\rm int}(t)-\mathcal{I}_{\rm int}(0)
 = \mathcal{I}_{\rm ext}(0)-\mathcal{I}_{\rm ext}(t)
 \leq \mathcal{I}_{\rm ext}(0),
\end{equation}
where $\mathcal{I}_{\rm ext}(t)$ is the information outside the open system 
defined in Eq.~\eqref{eq:1bis}. This inequality, as simple as it is, reveals one very 
important point: The information in the open system $\mathcal{I}_{\rm int}(t)$ can 
increase above its initial value $\mathcal{I}_{\rm int}(0)$ only if there is some 
information initially outside the system, i.e. $\mathcal{I}_{\rm ext}(0)>0$. 
Obviously, the contraction property \eqref{contraction-property} for completely 
positive maps is a special case of the inequality \eqref{initial-inequality-1} which 
occurs when the system and the environment are initially uncorrelated and the 
environmental state is not influenced by the system state preparation, i.e. 
$\rho_{SE}^{1,2} = \rho_S^{1,2} \otimes \rho_E$. This leads to 
$\mathcal{I}_{\rm ext}(0)=0$ and, thus, Eq.~\eqref{initial-inequality-1} reduces to
Eq.~\eqref{decrease}.

\subsubsection{Local detection of initial correlations}
\label{local-detection-scheme}

We found that initial information outside the open system can
lead to an increase of trace distance. But how can this be used to
develop experimental methods to detect correlations in some unknown
initial state $\rho^1_{SE}$? To this end, let us combine
\eqref{initial-inequality-1} and the inequality \eqref{bound-I-out} for the initial time 
in order to reveal the role of initial correlations more explicitly \cite{Laine2010b}:
\begin{eqnarray}\label{initial-inequality-2}
 \mathcal{I}_{\rm int}(t)-\mathcal{I}_{\rm int}(0)&\leq& D(\rho_{SE}^1(0),\rho_S^1(0)\otimes
 \rho_E^1(0))\nonumber\\
 &&+D(\rho_{SE}^2(0),\rho_S^2(0)\otimes\rho_E^2(0))\nonumber\\
 &&+D(\rho_E^1(0),\rho_E^2(0)). 
\end{eqnarray}
This inequality clearly shows that an increase of the trace distance of the reduced 
states implies that there are initial correlations in $\rho_{SE}^1(0)$ or 
$\rho_{SE}^2(0)$, or that the initial environmental states are different. 

\begin{figure}[tbh]
\includegraphics[width=0.40\textwidth]{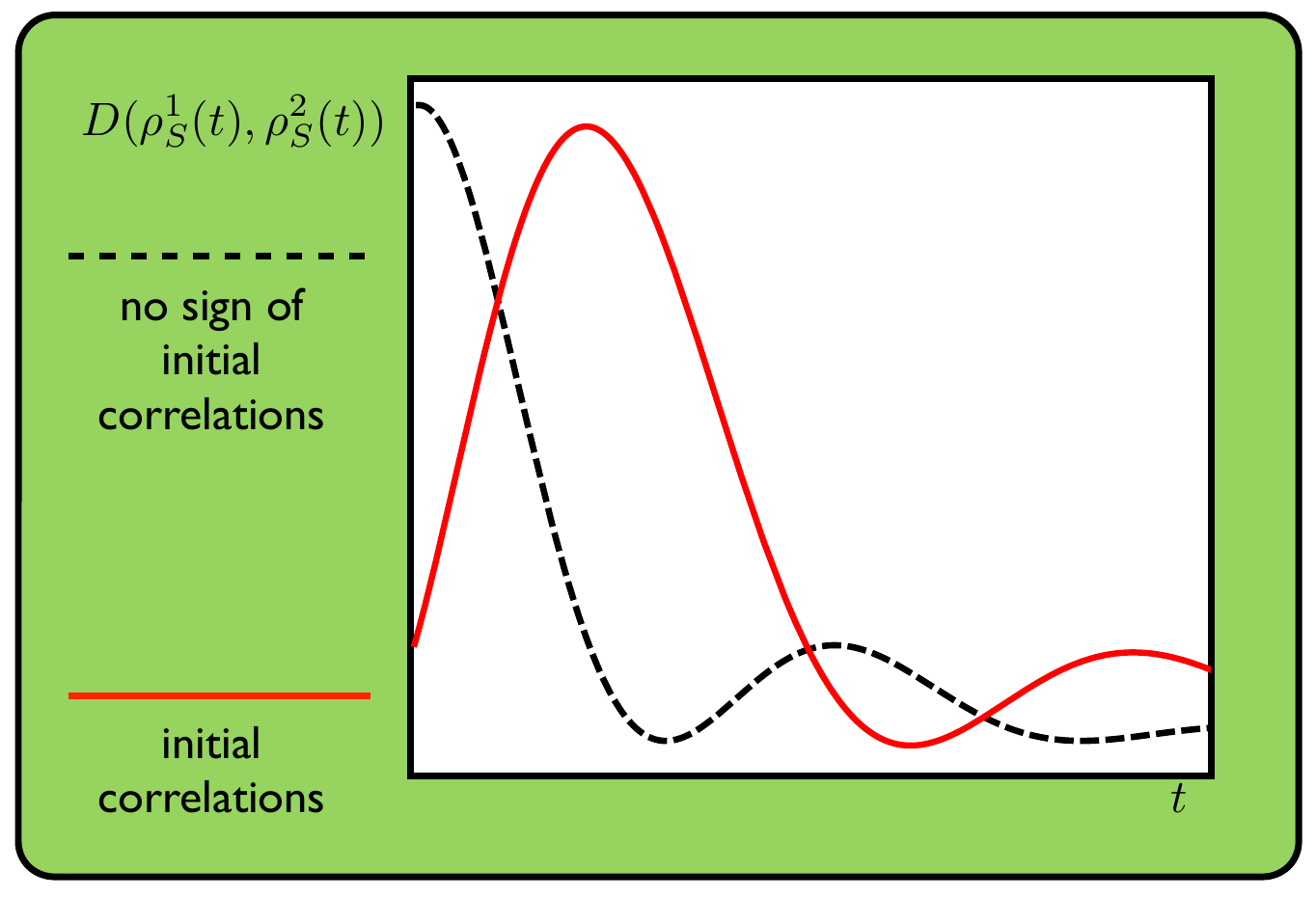}
\caption{(Color online) Schematic picture of the behavior of the trace distance with 
and without initial correlations. If the trace distance between two states, prepared 
such that $\rho_{SE}^2(0)=(\Lambda\otimes I)\rho_{SE}^1(0)$, never 
exceeds its initial value (dashed black line), the dynamics does not witness initial 
correlations. If, on the other hand, the trace distance for such pair of states 
increases above the initial value (solid red line), we can conclude the presence of 
initial correlations.}
\label{fig:init_corr}
\end{figure}

Let us now assume that one can perform a state tomography on the open system 
at the initial time zero and at some later time $t$, to determine the reduced states 
$\rho_S^1(0)$ and $\rho_S^1(t)$. In order to apply inequality 
\eqref{initial-inequality-2} to detect initial correlations we need a second reference state $\rho_S^2(0)$, 
which has the same environmental state, i.e. $\rho_E^2(0)=\rho_E^1(0)$. This 
can be achieved by performing a local quantum operation on $\rho_{SE}^1(0)$  to 
obtain the state 
\begin{equation}\label{second-reference-state}
 \rho_{SE}^2(0)=(\Lambda\otimes I)\rho_{SE}^1(0).
\end{equation}
The operation $\Lambda$ acts locally on the variables of the open system, and 
may be realized, for instance, by the measurement of an observable of the open 
system, or by a unitary transformation induced, e.g., through an external control 
field. Now, if $\rho_{SE}^1(0)$ is uncorrelated then also $\rho_{SE}^2(0)$ is 
uncorrelated since it has been obtained through a local operation. Therefore, for 
an initially uncorrelated state the trace distance cannot increase according to 
inequality \eqref{initial-inequality-2}. Thus, any such increase represents a witness 
for correlations in the initial state $\rho_{SE}^1(0)$, as is illustrated
in Fig.~\ref{fig:init_corr}.
We note that this method for the local detection of initial correlations 
requires only local control and measurements of the open quantum system, which 
makes it feasible experimentally and thus attractive for applications. In fact, 
experimental realizations of the scheme have been reported recently, see
Sec.~\ref{exp-local-detection}. Furthermore, \textcite{Smirne2010c} have studied 
how the scheme can be employed for detecting  correlations in thermal 
equilibrium states. 

\begin{figure}[tbh]
\includegraphics[width=0.40\textwidth]{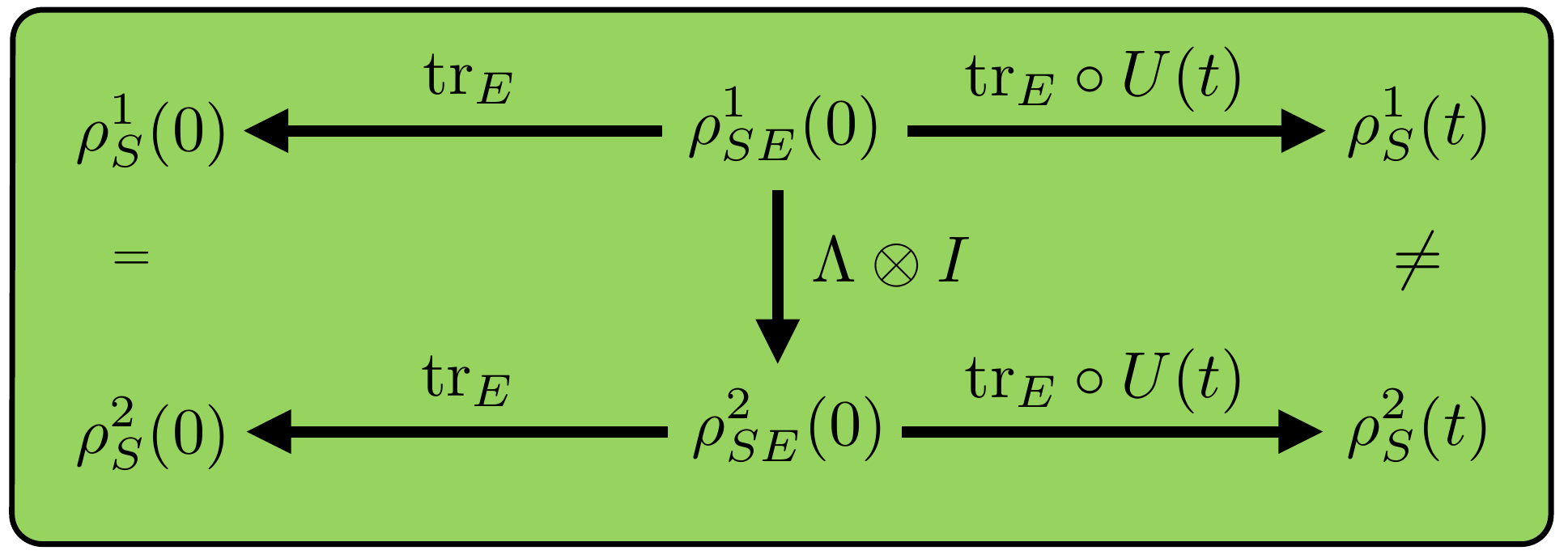}
\caption{(Color online) Scheme for the local detection of 
system-environment quantum correlations in some state $\rho^1_{SE}(0)$, 
employing a local dephasing operation $\Lambda$ to generate a second reference 
state $\rho^2_{SE}(0)$.}
\label{fig:local_detect}
\end{figure}

The above strategy can even be used in order to locally detect a specific 
type of quantum correlations, i.e. system-environment states with 
nonzero quantum discord \cite{Modi2012}. This is achieved by taking
the local operation $\Lambda$ in Eq.~\eqref{second-reference-state} to be a
dephasing operation leading to complete decoherence in the eigenbasis of
the state $\rho_S^1(0)$ \cite{Gessner2011a, Gessner2013a}.
The application of this dephasing operation destroys all quantum correlations of
$\rho^1_{SE}(0)$, while leaving invariant its marginal states $\rho^1_S(0)$ and
$\rho^1_E(0)$ (see Fig.~\ref{fig:local_detect}). Thus, one can use the quantity
${\mathcal{C}}(\rho^1_{SE}(0)) = D(\rho^1_{SE}(0),\rho^2_{SE}(0))$
as a measure for the quantum correlations in the initial state $\rho^1_{SE}(0)$
\cite{Luo2008}. Since the trace distance is invariant under unitary transformations 
and since the partial trace is a positive map, the corresponding local trace
distance $D(\rho^1_S(t),\rho^2_S(t))$ provides a lower bound of
${\mathcal{C}}(\rho^1_{SE}(0))$ for all times $t\geq 0$. Hence, we have
\begin{equation}\label{lower-bound-discord}
 \max_{t\geq 0} D(\rho^1_S(t),\rho^2_S(t)) \leq {\mathcal{C}}(\rho^1_{SE}(0)).
\end{equation}
The left-hand side of this inequality yields a locally accessible lower bound
for a measure of the quantum discord. Experimental realizations of the scheme
are briefly described in Sec.~\ref{exp-local-detection}.
Most recent studies suggest that the method could also be applied for 
detecting a critical point of a quantum phase transition via dynamical monitoring 
of a single spin alone \cite{Gessner2014b}.

\subsection{Nonlocal memory effects}\label{non_local}
We now return to study the case of initially uncorrelated system-environment 
states, for which the usual description in terms of dynamical maps can be used. 
So far, we have concentrated on examples with a single system embedded in an 
environment. However, many relevant examples in quantum information 
processing include multipartite systems for which nonlocal properties such as 
entanglement become important and it is therefore essential to study how 
non-Markovianity is influenced by scaling up the number of particles.

The question of additivity of memory effects with respect to particle
number was recently raised by \textcite{Addis2013a, Addis2014a,
Fanchini2013a} and it was found that the different measures for
non-Markovianity have very distinct additivity properties. Indeed, the
research on multipartite open quantum systems is yet in its infancy
and a conclusive analysis remains undone. In the following, we will
discuss a particular feature of bipartite open systems: the appearance
of global memory effects in the absence of local non-Markovian
dynamics. This is at variance with the standard situation in which the
enlargement of considered degrees of freedom leads from a
non-Markovian to a Markovian dynamics \cite{Martinazzo2011a}.

We will now take a closer look at nonlocal maps, for which memory
effects may occur even in the absence of local non-Markovian
effects. In \textcite{Laine2012a, Laine2013b} it is shown that such
maps may be generated from a local interaction, when correlations
between the environments are present and that they may exhibit
dynamics with strong global memory effects although the local dynamics
is Markovian. The nonlocal memory effects are studied in two dephasing models: 
in a generic model of qubits interacting with correlated multimode fields 
\cite{Wissmann2013b,Wissmann2014a} and in an experimentally realizable 
model of  down converted photons traveling through quartz plates, which we will 
discuss later in detail.

\begin{figure}[tbh]
\includegraphics[width=0.40\textwidth]{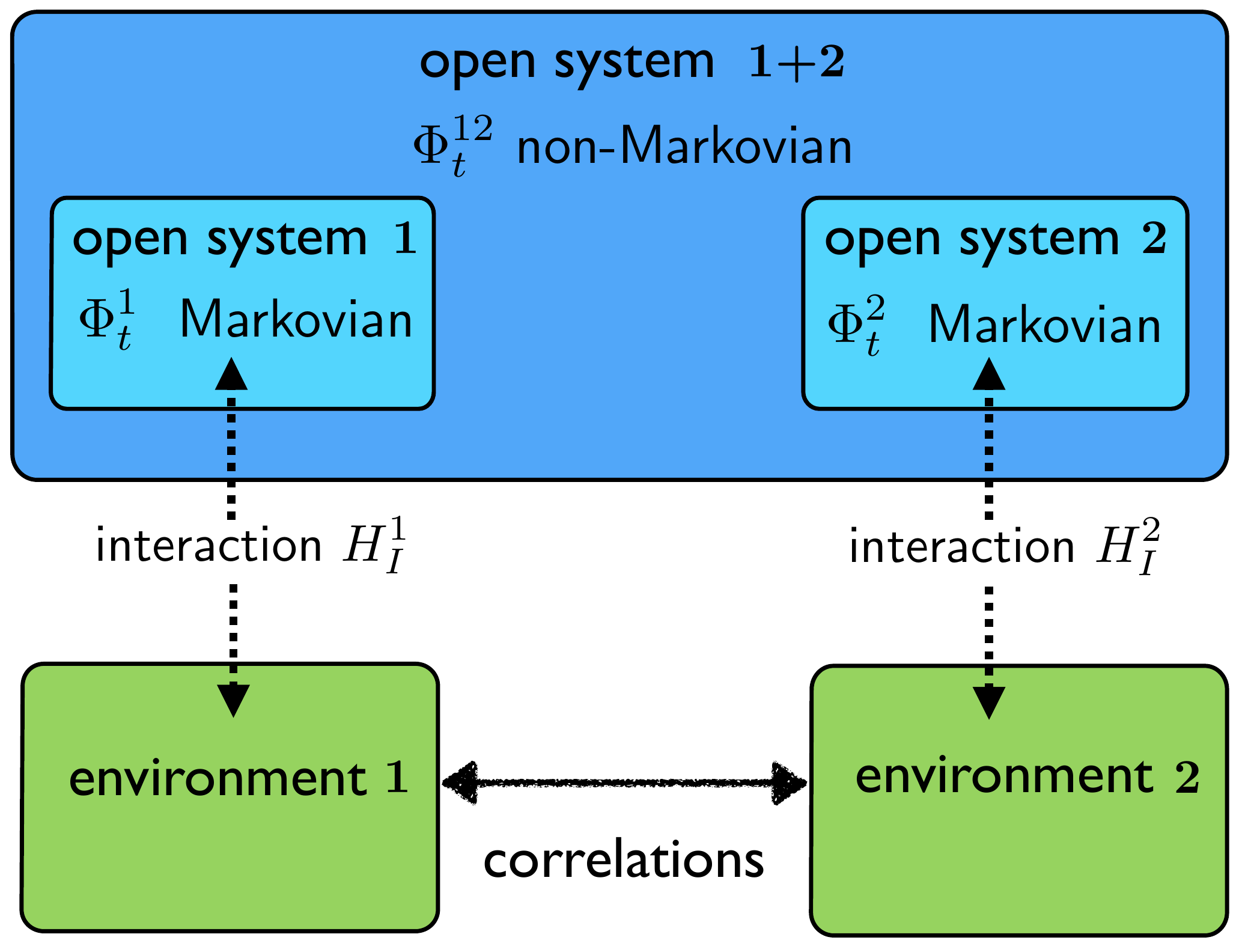}
\caption{(Color online) Schematic picture of a system with nonlocal memory 
effects. The open systems $1$ and $2$ locally interact with their respective environments. 
Initial correlations between the local environments cause the occurrence of 
nonlocal memory effects.}
\label{fig:nl_nm}
\end{figure}

Before we turn to discuss any experimental endeavor to detect non-Markovian 
dynamics, let us discuss in more rigor about the generation and possible 
applications of nonlocal memory effects. Consider a generic scenario, where 
there are two systems, labeled with indices $i=1,2$, which interact locally with 
their respective environments. The dynamics of the two systems can be described 
via the dynamical map
\begin{eqnarray}
 \rho_S^{12}(t) &=& \Phi_t^{12} \rho_S^{12}(0) \nonumber \\
 &=& \mathrm{tr}_E \left[
 U_{SE}^{12}(t) \rho_S^{12}(0)\otimes \rho_E^{12}(0) U_{SE}^{12\dagger}(t)
 \right],
 \label{nonlocal_map}
\end{eqnarray}
where $U_{SE}^{12}(t) = U_{SE}^{1}(t) \otimes U_{SE}^{2}(t)$
with $U_{SE}^i(t)$ describing the local interaction between the system $i$ 
and its environment. If the initial environment state 
$\rho_E^{12}(0)$ factorizes, i.e.  
$\rho_E^{12}(0)=\rho_E^1(0)\otimes\rho_E^2(0)$, also the map $\Phi_t^{12}$ 
factorizes. Thus, the dynamics of the two systems is given by a local map 
$\Phi_t^{12}=\Phi_t^1\otimes\Phi_t^2$. On the other hand, if $\rho_E^{12}(0)$ 
exhibits correlations, the map $\Phi_t^{12}$ cannot, in general, be factorized. 
Consequently, the environmental correlations may give rise to a nonlocal process 
even though the interaction Hamiltonian is purely local (see Fig.~\ref{fig:nl_nm}).

For a local dynamical process, all the dynamical properties of the subsystems are 
inherited by the global system, but naturally for a nonlocal process the global 
dynamics can display characteristics absent in the dynamics of the local 
constituents. Especially, for a nonlocal process, even if the subsystems undergo a 
Markovian evolution, the global dynamics can nevertheless be highly 
non-Markovian as was shown for the dephasing models by
\textcite{Laine2012a, Laine2013b}.

Thus, a system can globally recover its earlier lost quantum properties although 
the constituent parts are undergoing decoherence and this way initial 
environmental correlations can diminish the otherwise destructive effects of 
decoherence. This feature has been further deployed in noisy quantum 
information protocols, such as teleportation \cite{Laine2014a} and entanglement 
distribution \cite{Xiang2014a}, suggesting that non-Markovianity could be a 
resource for quantum information.

\subsection{Experiments on non-Markovianity and correlations}
Up to this point we have not yet discussed any experimental aspects of the 
detection of memory effects. In the framework of open quantum systems the 
environment is in general composed of many degrees of freedom and is therefore 
difficult to access or control. To perform experiments on systems where the 
environment induced dynamical features can be controlled is thus challenging. 
However, in the past years clever schemes for modifying the environment have 
been developed allowing the establishment of robust designs for noise 
engineering. In this section we will briefly review some of the experimental 
platforms where a high level of control over the environment degrees of freedom 
has been accomplished and non-Markovian dynamics observed and quantified.

\subsubsection{Control and quantification of memory effects in photonic systems}
\label{sec:contr-quant-memory}
Quantum optical experiments have for many decades been the bedrock for 
testing fundamental paradigms of quantum mechanics. The appeal for using 
photonic systems arises from the extremely high level of control allowing, for 
example, controlled interactions between different degrees of freedom, 
preparation of arbitrary polarization states and a full state tomography. Needless 
to say, photons thus offer an attractive experimental platform also for studying 
non-Markovian effects.

\textcite{Liu2011a} introduced an all-optical experiment which allows through 
careful manipulation of the initial environmental states to drive the open system 
dynamics from the Markovian to the non-Markovian regime, to control the 
information flow between the system and the environment, and to determine the 
degree of non-Markovianity. In the experiment the photon polarization degree of 
freedom (with basis states $\ket{H}$ and $\ket{V}$ for horizontal
and vertical polarization, respectively) plays the role of the open 
system. The environment is represented by the frequency degree of freedom
of the photon (with basis $\left\{\ket{\omega}\right\}_{\omega\geq 0}$), 
which is coupled to the system 
via an interaction induced by a birefringent material (quartz plate). The interaction 
between the polarization and frequency degrees of freedom in the quartz plate of 
thickness $L$ is described by the unitary operator 
$U(t) \ket{\lambda}\otimes\ket{\omega}=e^{i n_\lambda \omega t}\ket{\lambda}
\otimes\ket{\omega}$, where $n_\lambda$ is the refraction index for a photon with 
polarization $\lambda=H,V$ and $t=L/c$ is the interaction time with the 
speed of light $c$. The photon is 
initially prepared in the state $\ket{\psi}\otimes\ket{\chi}$, with 
$\ket{\psi}=\alpha\ket{H}+\beta\ket{V}$ and $\ket{\chi}=\int d\omega f(\omega) 
\ket{\omega}$, where $f(\omega)$ gives the amplitude for the photon to be in a 
mode with frequency $\omega$. The quartz plate leads to a pure decoherence 
dynamics of superpositions of polarization states (see Sec.~\ref{sec:dephasing})
described by the complex decoherence function 
$G(t)=\int d\omega |f(\omega)|^2 e^{i \omega \Delta n t}$,
where $\Delta n=n_V-n_H$. An optimal pair of states maximizing the non-Markovianity 
measure \eqref{NM-measure} for this map is given by
$\ket{\psi_{1,2}}=\frac{1}{\sqrt{2}}(\ket{H}\pm\ket{V})$ and the corresponding 
trace distance is $D(\rho_S^1(t),\rho_S^2(t))=|G(t)|$. Clearly, controlling the 
frequency spectrum $|f(\omega)|^2$ changes the dynamical features of the map.

\begin{figure}[tbh]
\includegraphics[width=0.35\textwidth]{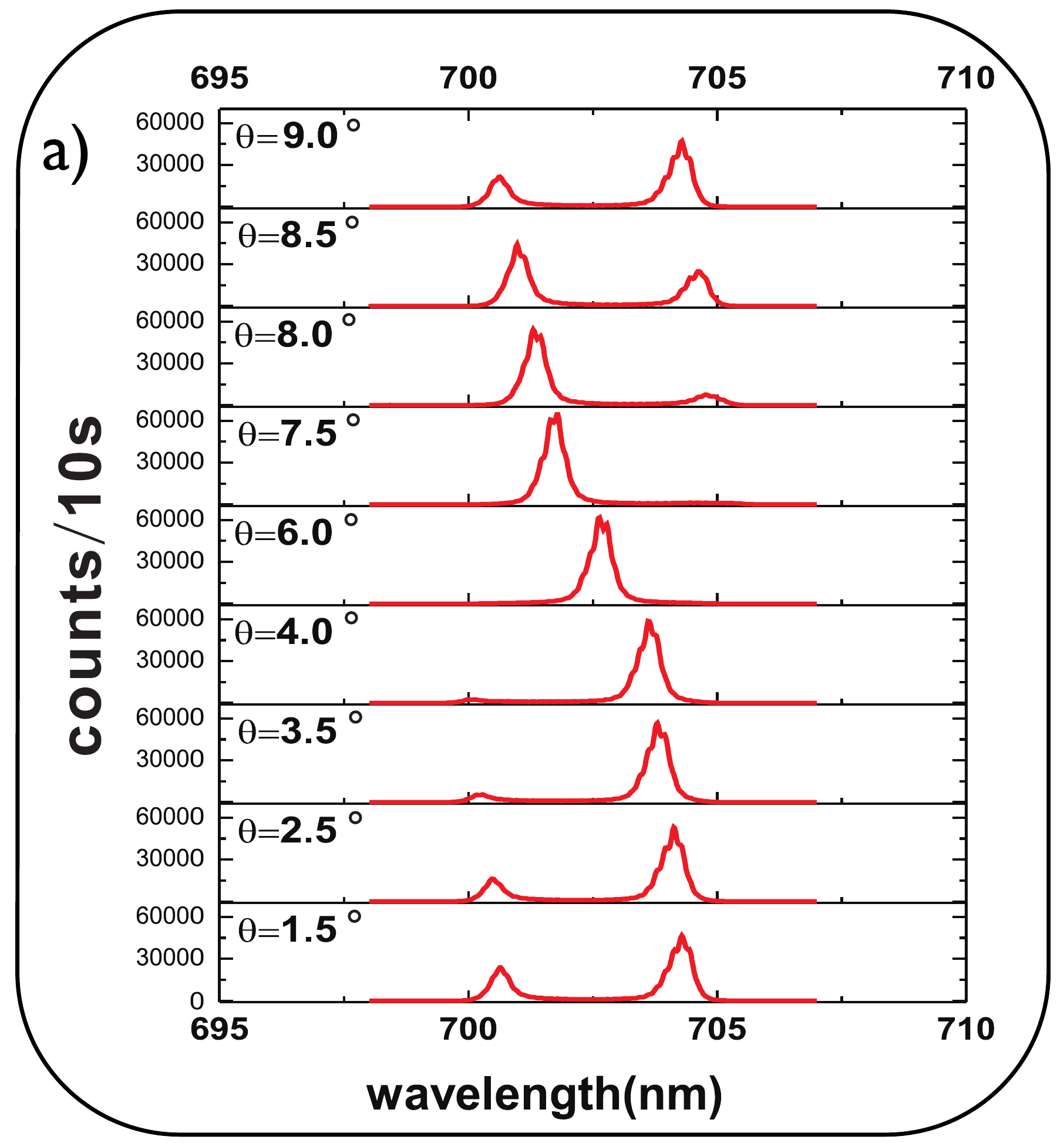}
\includegraphics[width=0.35\textwidth]{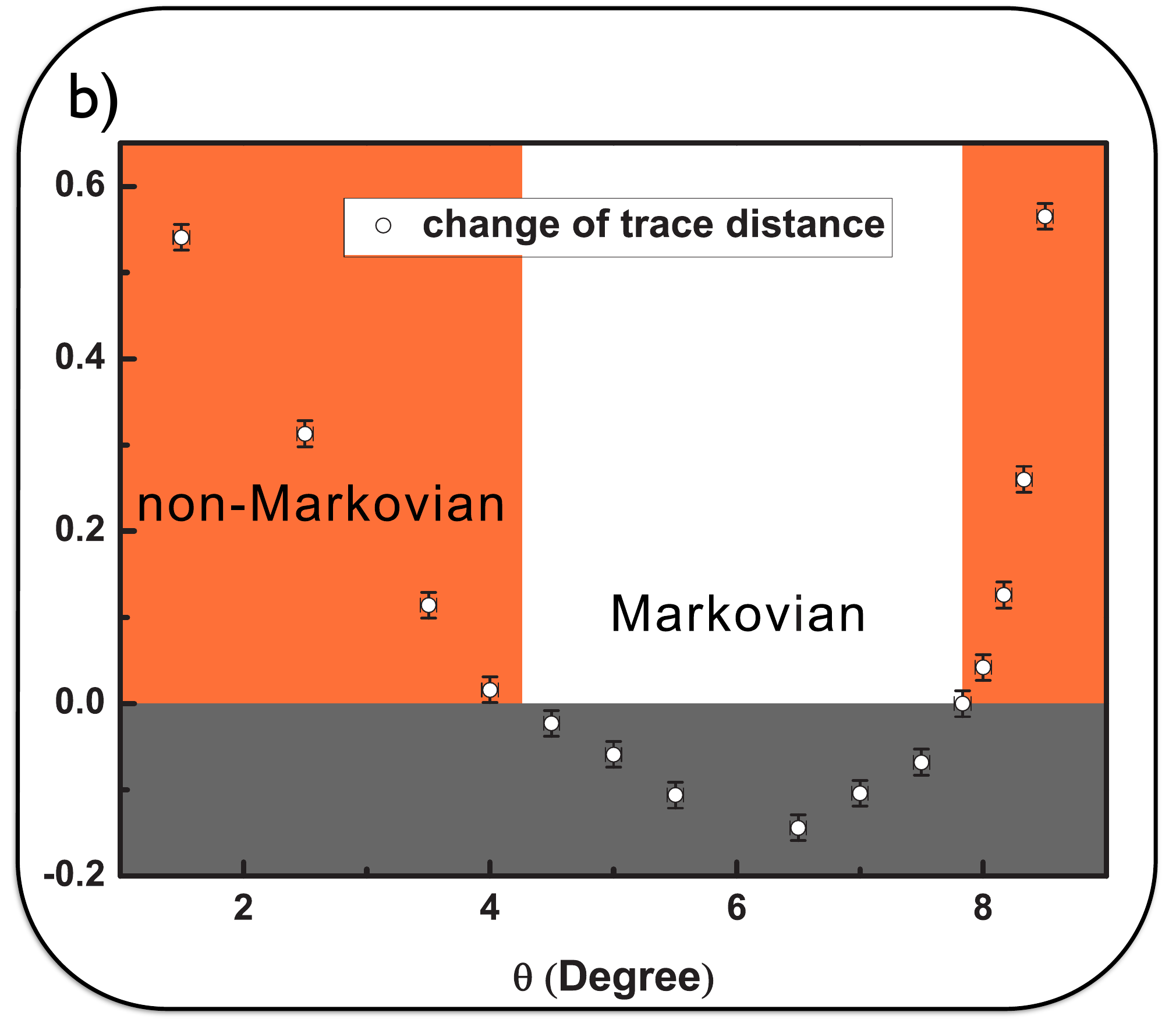}
\caption{(Color online) Non-Markovian dynamics arising from an engineered 
frequency spectrum. (a) The frequency spectrum of the initial state for various 
values of the tilting angle $\theta$ of the cavity. (b) The change of the trace 
distance as functions of the tilting angle $\theta$. The transition from the 
non-Markovian to the Markovian regime occurs at $\theta \approx 4.1^{\circ}$, 
and from the Markovian to the non-Markovian regime at 
$\theta \approx 8.0^{\circ}$, corresponding to the occurrence of a double peak 
structure in the frequency spectrum (a) \cite{Liu2011a}.}
\label{fig:transition_mnm}
\end{figure}

The modification of the photon frequency spectrum is realized by means of a 
Fabry-P\'{e}rot cavity 
mounted on a rotator which can be tilted in the horizontal plane. The shape of the 
frequency spectrum is changed by changing the tilting angle $\theta$ of the Fabry-P\'{e}rot 
cavity, as can be seen in Fig.~\ref{fig:transition_mnm}a. Further, the frequency 
spectrum changes the dephasing dynamics such that the open system exhibits a 
reversed flow of information and thus allows to tune the dynamics from Markovian 
to non-Markovian regime (see  Fig.~\ref{fig:transition_mnm}b). In the experiment, 
full state tomography can further be performed thus allowing a rigorous 
quantification of the memory effects.

Also, a series of other experimental studies on non-Markovian dynamics in 
photonic systems have been performed recently. \textcite{Tang2012a} report a 
measurement of the non-Markovianity of a process with tunable system-
environment interaction,  \textcite{Cialdi2011a} observe controllable entanglement 
oscillations in an effective non-Markovian channel and in 
\cite{Chiuri2012a, Jin2014a}  simulation platforms for a wide class of 
non-Markovian channels are presented.

\subsubsection{Experiments on the local detection of correlations}
\label{exp-local-detection}

A number of experiments have been carried out in order to demonstrate the local
scheme for the detection of correlations between an open system and its 
environment described in Sec.~\ref{local-detection-scheme}. 
Photonic realizations of this scheme have been reported in 
\cite{Li2011a,Smirne2011a}, where the presence of initial correlations 
between the polarization and the spatial degrees of freedom of photons 
is shown by the observation of an increase of the local trace distance between a 
pair of initial state. 

As explained in Sec.~\ref{local-detection-scheme}, it is also possible to
reveal locally the presence of quantum correlations represented by 
system-environment states with nonzero discord if the local operation 
$\Lambda$ in Eq.~\eqref{second-reference-state} is taken to induce 
complete decoherence in the eigenbasis 
of the open system state. The first photonic realizations of this strategy 
based on Eq.~\eqref{lower-bound-discord} is described in \cite{Tang2014}. 
Moreover, \textcite{Cialdi2014a} have extended the method to enable the
discrimination between quantum and classical correlations, and applied
this extension to a photonic experimental realization.

All experiments mentioned so far employ photonic degrees of freedom
to demonstrate non-Markovianity and system-environment correlations. 
The first experiment showing these phenomena for matter degrees of freedom has been 
described by \textcite{Gessner2014a}. In this experiment nonclassical correlations 
between the internal electronic degrees of freedom and the external motional 
degrees of freedom of a trapped ion have been observed and quantified by use
of Eq.~\eqref{lower-bound-discord}.
Important features of the experiment are that the lower bounds obtained from the
experimental data are remarkably close to the true quantum correlations 
present in the initial  state, and that it also allows the study of the temperature 
dependence of the effect.

\subsubsection{Non-Markovian quantum probes detecting nonlocal correlations in 
composite environments}
In Sec.~\ref{non_local} we demonstrated that initial correlations between local 
parts of the environment can lead to nonlocal memory effects. In 
\textcite{Liu2013a} such nonlocal memory effects were experimentally realized in 
a photonics system, where manipulating the correlations of the photonic 
environments led to non-Markovian dynamics of the open system. The 
experimental scheme further provided a controllable diagnostic tool for the 
quantification of these correlations by repeated tomographic measurements of the 
polarization.

Let us take a closer look at the system under study in the experiment. A general 
pure initial polarization state of a photon pair can be written as 
$\ket{\psi_{12}}=a \ket{HH}+b\ket{HV}+c\ket{VH}+d\ket{VV}$ and all initial 
states of the polarization plus the frequency degrees of freedom are product 
states
\begin{equation} \label{eq51}
 \ket{\Psi(0)} = \ket{\psi_{12}} \otimes \int d\omega_1 d\omega_2 \,
 g(\omega_1,\omega_2) \ket{\omega_1,\omega_2},
\end{equation}
where $g(\omega_1,\omega_2)$ is the probability amplitude for photon $1$ to 
have frequency $\omega_1$ and for photon $2$ to have frequency 
$\omega_2$, with the corresponding joint probability distribution 
$P(\omega_1,\omega_2)=|g(\omega_1,\omega_2)|^2$. If the photons pass 
through quartz plates, the dynamics can be described by a general two-qubit 
dephasing map, where the different decoherence functions can be expressed in 
terms of Fourier transforms of the joint probability distribution 
$P(\omega_1,\omega_2)$. For a Gaussian joint frequency distribution with 
identical single frequency variances $C$ and correlation coefficient $K$, the time 
evolution of the trace distance corresponding to the Bell-state pair 
$\ket{\psi^{\pm}_{12}}=\frac{1}{\sqrt{2}}(\ket{HH}\pm \ket{VV})$ is found to be
\begin{equation} 
 D(t)= \exp\left[
 -\frac{1}{2}\Delta n^2C \left( t_1^2 + t_2^2 - 2|K|t_1t_2 \right) \right],
\end{equation} 
where $t_i$ denotes the time photon $i$ has interacted with its quartz plate
up to the actual observation time $t$.
For uncorrelated photon frequencies we have $K=0$ and the trace distance 
decreases monotonically, corresponding to Markovian dynamics. However, as 
soon as the frequencies are anticorrelated, $K<0$, the trace distance is non-
monotonic which signifies quantum memory effects and non-Markovian behavior. 
On the other hand, the local frequency distributions are Gaussian and thus for the 
single photons the trace distance monotonically decreases. Therefore we can 
conclude, that the system is locally Markovian but globally displays nonlocal 
memory effects.

In the experiment two quartz plates act consecutively for the photons, and the 
magnitude of the initial anticorrelations between the local reservoirs is tuned. After 
the photon exits the quartz plates, full two-photon polarization state tomography is 
performed and by changing the quartz plate thicknesses, the trace distance 
dynamics is recovered. From the dynamics it is evident that initial environmental 
correlations influence the quantum non-Markovianity. 

A further important aspect of the experimental scheme is that it enables to 
determine the frequency correlation coefficient $K$ of the photon pairs from 
measurements performed on the polarization degree of freedom. Thus by 
performing tomography on a small system we can obtain information on 
frequency correlations, difficult to measure directly. Thus, in this context, the open 
system (polarization degrees of freedom) can serve as a quantum probe which 
allows us to gain nontrivial information on the correlations in the environment 
(frequency degrees of freedom).


\section{Summary and outlook}

In this Colloquium we have presented recent advances in the definition
and characterization of non-Markovian dynamics for an open quantum
system.  While in the classical case the very definition of Markovian
stochastic process can be explicitly given in terms of constraints on
the conditional probabilities of the process, in the quantum realm the
peculiar role of measurements prevents a direct formulation along the
same path and new approaches are called for in order to describe
memory effects. We have therefore introduced a notion of
non-Markovianity of a quantum dynamical map based on the behavior in
time of the distinguishability of different system states, as
quantified by their trace distance, which provides an intrinsic
characterization of the dynamics. This approach allows to connect
non-Markovianity with the flow of information from the environment back
to the system and naturally leads to the introduction of quantum
information concepts. It is further shown to be connected to other
approaches recently presented in the literature building on
divisibility properties of the quantum dynamical map. In particular, a
generalization of the trace distance criterion allows to identify
Markovian time evolutions with quantum evolutions which are P-divisible,
thus leading to a clear-cut connection between memory effects in the
classical and quantum regimes. As a crucial feature the trace
distance approach to non-Markovianity can be experimentally tested and
allows for the study of system-environment correlations, as well as nonlocal 
memory effects emerging from correlations within the environment.

We have illustrated the basic feature of the considered theoretical
approaches to quantum non-Markovianity, typically leading to a
recovery of quantum coherence properties, by analyzing in detail
simple paradigmatic model systems, and further discussing more complex
models which show important connections between non-Markovianity of
the open system dynamics and features of the environment. It is indeed
possible to obtain information on complex quantum systems via study of
the dynamics of a small quantum probe.  The study of the time
development of the trace distance between system states can be shown
to provide an indication of the presence of initial correlations
between system and environment, thus providing a powerful tool for the
local detection of correlations. Further considering suitable local
operations, such correlations can be further characterized and, in
particular, one can distinguish quantum and classical correlations.

Experimentally, it is in general obviously very difficult to control
in detail the environmental degrees of freedom and therefore experiments
on non-Markovian quantum systems are still in a state of
infancy. Still, some sophisticated schemes for modifying the
environment have been developed, thus leading to the realization of
first proof-of-principle experiments in which to test the study of
non-Markovianity, its connection with crucial features of the
environment, as well as the capability to unveil system-environment
correlations by means of local observations on the system.

In recent years non-Markovian quantum systems have been enjoying much
attention both due to fundamental reasons and foreseeable applications,
as shown by the rapid growth in the related literature.
Indeed, the potential relevance of memory effects in the field of
complex quantum systems and quantum information has led to an intense
study, and we have pointed to some highlights in this novel, so far
fairly unexplored field of non-Markovianity. However, there are numerous open questions
yet to be studied. The latter include fundamental questions 
such as the mathematical structure of the space of non-Markovian quantum dynamical
maps, the role of complexity in the emergence of memory effects or the relevance of 
non-Markovianity in the study of the border between
classical and quantum aspects of nature, as well as more
applied issues like the identification of the environmental features
or system-environment correlations which
can indeed be detected by means of local observations on the system.

The theoretical and experimental investigations of non-Markovian quantum systems 
outlined in this Colloquium pave the way for new lines of research
by both shedding light on fundamental questions of open quantum 
systems as well as by suggesting novel applications in quantum information 
and probing of complex systems.

\begin{acknowledgments}
HPB, JP and BV acknowledge support from the
EU Collaborative Project QuProCS (Grant Agreement 641277).
JP and BV acknowledge support by the COST MP1006 Fundamental Problems 
in Quantum Physics, EML by the Academy of Finland through its Centres of Excellence 
Programme (2012-2017) under project No. 251748,
JP by the Jenny and Antti Wihuri and by the Magnus Ehrnrooth 
Foundation, and BV by the Unimi TRANSITION GRANT - HORIZON 2020.
\end{acknowledgments}

\bibliography{biblio_nonmarkov}

\end{document}